\documentclass[reprint,aps,pra,twocolumn,amsmath,amssymb,superscriptaddress,floatfix,footinbib,longbibliography]{revtex4-2}

\usepackage{epsfig}
\usepackage{graphicx}
\usepackage{epstopdf}
\usepackage[T1]{fontenc}
\usepackage[utf8]{inputenc}
\usepackage{lmodern}
\usepackage[english]{babel}
\usepackage{lipsum}
\usepackage{ae}
\usepackage{siunitx}
\usepackage{dsfont}

\usepackage{natbib}
\usepackage{amsmath,amssymb,bm}
\usepackage{psfrag}
\usepackage{subfigure}
\usepackage{amsthm}
\usepackage{bbm}
\usepackage{booktabs}

\usepackage{tikz}
\usetikzlibrary{arrows}

\usepackage{color}
\usepackage{url}
\usepackage[colorlinks]{hyperref}
\hypersetup{%
        plainpages=true,
        breaklinks=true,
        hypertexnames=false,
        pageanchor=true,
        colorlinks=true,
        linkcolor={blue},
        citecolor={magenta},
        urlcolor={blue},
        anchorcolor={black}
      }
\usepackage[all]{hypcap} 

\usepackage{mleftright} 

\renewcommand{\eqref}[1]{\mbox{Eq.~(\ref{#1})}}

\newcommand{\figpanel}[2]{Fig.~\hyperref[#1]{\ref*{#1}(#2)}}
\newcommand{\figpanels}[3]{Fig.~\hyperref[#1]{\ref*{#1}(#2)-(#3)}}
\newcommand{\figpanelNoPrefix}[2]{\hyperref[#1]{\ref*{#1}(#2)}}

\newcommand{\brakket}[3]{\mleft\langle #1\mleft| #2 \mright| #3\mright\rangle}
\newcommand{\brakketsmall}[3]{\langle #1| #2 | #3\rangle}

\newcommand{\comm}[2]{\mleft[ #1, #2 \mright]}

\newcommand{\abs}[1]{\mleft|#1\mright|}

\newcommand{\be}{\begin{equation}}
\newcommand{\ee}{\end{equation}}
\newcommand{\bea}{\begin{eqnarray}}
\newcommand{\eea}{\end{eqnarray}}

\DeclareMathOperator{\Tr}{Tr} 

\newcommand{\ket}[1]{\mleft|#1\mright\rangle}
\newcommand{\bra}[1]{\mleft\langle#1\mright|}
\newcommand{\braket}[2]{\langle #1|#2\rangle}
\newcommand{\ketbra}[2]{\mleft| #1 \rangle \langle #2 \mright|}

\renewcommand*{\i}{\mathrm{i}}      
\newcommand*{\e}{\mathrm{e}}        
\renewcommand*{\d}{\mathrm{d}}      

\begin{document}

\title{Fidelity bounds for adiabatic gates and other quantum operations \\with time-dependent dissipation}




\author{Simon Pettersson Fors}
\affiliation{Department of Microtechnology and Nanoscience, Chalmers University of Technology, 412 96 Gothenburg, Sweden}

\author{Aniket Patel}
\affiliation{Department of Microtechnology and Nanoscience, Chalmers University of Technology, 412 96 Gothenburg, Sweden}

\author{Anton Frisk Kockum}
\affiliation{Department of Microtechnology and Nanoscience, Chalmers University of Technology, 412 96 Gothenburg, Sweden}

\author{Tahereh Abad}
\email{tahereh.abad@chalmers.se}
\affiliation{Department of Microtechnology and Nanoscience, Chalmers University of Technology, 412 96 Gothenburg, Sweden}

\begin{abstract}

As quantum-computing platforms are susceptible to noise, the fidelity of quantum operations is limited by decoherence. Understanding this limitation is crucial for building utility-scale quantum processors. In previous works [\href{https://doi.org/10.1103/PhysRevLett.129.150504}{Phys.~Rev.~Lett.~\textbf{129}, 150504 (2022)}; \href{https://doi.org/10.22331/q-2025-04-03-1684}{Quantum \textbf{9}, 1684 (2025)}], we presented analytical formulae for the average gate fidelity of multi-qubit operations under static Markovian noise processes, including operations that temporarily leave the computational subspace. However, some quantum-computing architectures dynamically modulate qubit or coupler frequencies to implement two-qubit gates, e.g., baseband flux gates; such modulation can lead to dissipation rates varying in time. In this Letter, we therefore generalize the fidelity-reduction formulae to encompass time-dependent dissipation. Applying our generalized formula, we obtain a fidelity bound for adiabatic operations and demonstrate that flux-dependent noise sensitivity, combined with qubit-coupler hybridization, significantly reduces the fidelity of adiabatic controlled-Z (CZ) gates in superconducting quantum computers. Our work thus provides essential theoretical tools for evaluating error budgets and optimizing the design of quantum operations in tunable quantum-computing architectures, and may also find applications in quantum-sensing and quantum-communication protocols that are affected by time-dependent dissipation.

\end{abstract}

\maketitle

\paragraph*{Introduction.}

The realization of large-scale quantum computers requires exquisite control over fragile quantum states~\cite{Nielsen2000, Mohseni2025}. Over the past decades, remarkable experimental progress has been achieved across a wide variety of quantum-computing hardware platforms, including superconducting circuits~\cite{Wendin2017, Krantz2019, Acharya2025}, trapped ions~\cite{Haffner2008, Bruzewicz2019, Strohm2024, Ransford2025}, neutral atoms~\cite{Saffman2010, Wintersperger2023, Bluvstein2026}, semiconductor spins~\cite{Chatterjee2021, Burkard2023, HRL2026}, photonic systems~\cite{Flamini2019, Slussarenko2019, AghaeeRad2025}, and rare-earth ions in solids~\cite{Kinos2021}. Despite these advancements, the inevitable coupling between the quantum processor and its surrounding environment introduces errors in the form of decoherence~\cite{Schlosshauer2019}, which degrades the performance of quantum gates. Mitigating this environment-induced loss is essential for reaching the operational thresholds demanded by quantum error correction for fault-tolerant quantum computing~\cite{Terhal2015, Girvin2023, Mohseni2025}. Beyond quantum computing, decoherence also imposes limits on the performance of quantum sensors~\cite{Degen2017, Montenegro2025, Khan2025} and quantum-communication protocols~\cite{Gisin2007, Hajdusek2023}. Therefore, precise understanding and quantification of open-system dynamics is a central pursuit in quantum information science for the development of quantum technologies.

The need for understanding the impact of decoherence has recently become more acute in quantum computers due to improvements in optimization of control signals for quantum gates. This control has now advanced to a level where decoherence, rather than coherent control infidelities, often is the predominant source of error for the gates~\cite{Sung2021, Ding2023, Li2024, Hyyppa2024}. Although analytical average-gate-fidelity formulae exist for errors due to static Markovian noise~\cite{Abad2022, Abad2025}, these static expressions break down for some tunable quantum-computing architectures, which are becoming increasingly common. 

For example, in superconducting circuits, baseband flux gates have emerged as a highly successful class of two-qubit operations~\cite{Neeley2010, Chen2014, Caldwell2018, Krantz2019, Barends2019, Sung2021, Ding2023, Li2024, Ding2025, Marxer2025, Chen2026}. By dynamically modulating qubit or coupler frequencies, these operations bring states into resonance or perform adiabatic controlled-Z (CZ) gates. In such cases, the noise sensitivity of the gate depends strongly on the instantaneous flux tuning of the qubit or coupler, rendering both the decoherence rates and the associated jump operators inherently time-dependent. Therefore, rigorously capturing the impact of decoherence on the fidelity of these operations becomes challenging. Although recent efforts provide insightful heuristic formulae to estimate decoherence errors in these dynamical scenarios~\cite{Li2024, Marxer2025, Chen2026}, a complete analytical treatment from first principles has so far been lacking.

In this Letter, we address this challenge by deriving an analytical formula for the fidelity reduction of quantum operations subject to time-dependent weak decoherence. Building upon the formalism used to obtain our previous results for static Markovian noise~\cite{Abad2022, Abad2025}, we generalize the average-gate-fidelity formula to accommodate dynamically varying dissipation rates. Applying this generalized result, we extract a rigorous fidelity bound for adiabatic operations. As a concrete example, we analyze adiabatic CZ gates, illustrating how the interplay between parameter-dependent noise channels and mode hybridization inherently restricts the achievable gate fidelity. Ultimately, our analytical expressions supply the physical insights necessary to reliably estimate error budgets and refine the control of dynamically tuned quantum hardware, both for quantum computing and for other quantum technologies.

\paragraph*{Average gate fidelity for time-dependent dissipation.}

To understand the effect of time-dependent dissipation on the fidelity of quantum operations, we begin from the definition of the average gate fidelity~\cite{Nielsen2002}
\be \label{fidelity_integral}
 \overline{F} \equiv \int \mathrm{d} \ket{\psi} \, \brakketsmall{\Psi}{\hat U_g^\dag \mathcal{E}( \ketbra{\Psi}{\Psi}) \hat U_g}{\Psi},
\ee
which measures the overlap between the state evolved by the (noisy) quantum channel $\mathcal{E}$ and the ideal unitary gate operation $\hat U_g$. The integral in \eqref{fidelity_integral} is taken over all pure initial states $\ket{\psi}$ in the computational subspace; the integrand varies with the initial state $\ket{\Psi} = \ket{\psi} \oplus \ket{\mathbf{0}}$, where $\ket{\mathbf{0}}$ is a zero vector in the relevant Hilbert space outside the computational subspace. Note that, as in Ref.~\cite{Abad2025}, the channel $\mathcal{E}$ can take states out of the computational subspace, and that $\mathcal{E}$ therefore does not necessarily preserve the trace in the computational subspace. For $\mathcal{E}$ perfectly implementing $\hat U_g$, we have $\overline{F}=1$. 

The gate operation in \eqref{fidelity_integral} can be generated by a time-dependent Hamiltonian $\hat H(t)$ applied for a time $\tau$, such that $\hat U_g = \hat U(\tau) = {\cal T} \exp[-\frac{i}{\hbar}\int_{0}^{\tau} \hat H(t) dt]$, where ${\cal T}$ is the time-ordering operator. This time dependence for the Hamiltonian was already considered in Refs.~\cite{Abad2022, Abad2025}, but there the noise terms were constant. Here, we extend the derivation in Refs.~\cite{Abad2022, Abad2025} by generalizing to the case where also the noise is time-dependent. Assuming $N_L$ different dissipative processes, each with a time-dependent rate $\Gamma_k(t)$ and a time-independent Lindblad jump operator $\hat{L}_k$, the time evolution of the system is then given by the master equation
\begin{equation} \label{master}
\dot{\hat \rho}(t) = -\frac{i}{\hbar} \comm{\hat H(t)}{\hat \rho(t)} + \sum_{k = 1}^{N_L} \Gamma_k(t) \mathcal{D} [\hat{L}_k] \hat \rho(t),
\end{equation}
where $\mathcal{D}[\hat L] \hat \rho = \hat L \hat \rho \hat L^\dagger - \frac{1}{2} \{\hat L^\dagger \hat L, \hat \rho \}$ is the standard Lindblad superoperator~\cite{Lindblad1976}. Note that we do not assume any restrictions preventing the ideal gate evolution or the jump operators from taking the system out of the computational subspace.

To describe the weakly dissipative dynamics of the system
, we expand the solution to the master equation in the small parameter $\bar{\Gamma}_{k} \tau \ll 1$, 
where $\bar{\Gamma}_{k} = \frac{1}{\tau} \int_0^\tau dt \, \Gamma_k(t)$ is the  time-averaged rate for the $k$th process.
We find from our previous results in Refs.~\cite{Abad2022, Abad2025}, that each dissipative process contributes independently to $\bar{F}$, and that those results directly generalize to first order in $\bar{\Gamma}_{k} \tau$~\cite{SuppMat}:
\be \label{fidelity}
\bar{F} = 1 + \sum_{k = 1}^{N_L} \int_0^\tau \mathrm{d} t \, \Gamma_k(t) \delta F[\hat{L}_k (t)] + \mathcal{O}\mleft(\bar{\Gamma}_k^2 \tau^2 \mright),
\ee
with the minor modification in the generalized case that $\Gamma_k(t)$ becomes a kernel in the time integral instead being a multiplicative constant in front of the integral. Here, the jump operators that represent the dissipative processes evolve in the \textit{Heisenberg picture} as
\be \label{L-time-dependent}
\hat{L}(t) = \hat U^\dag(t) \,\hat{L}\,\hat U(t).
\ee

The fidelity-reduction integrand for the $N$-qubit operation can be written as~\cite{Abad2025}
\bea
\label{deltaF}
\delta F(\hat{L}(t)) = \frac{1}{d(d+1)} \Tr_{\rm cmp}{\mleft[ \hat{L}^\dag(t) \mright]} \Tr_{\rm cmp}{\mleft[ \hat{L}(t) \mright]} \notag \\
 + \frac{1}{d(d+1)} \text{Tr}_{\rm cmp} \mleft[ \hat{L}^\dag(t) \hat{\mathds{1}}_{\rm cmp} \hat{L}(t) \hat{\mathds{1}}_{\rm cmp} \mright] \notag \\
 - \frac{1}{d}} \Tr_{\rm cmp}{\mleft[ \hat{L}^\dag(t) \, \hat{L}(t) \mright],
\eea 
where ``cmp'' denotes that the trace is over the states in the computational subspace. Here, $\hat{\mathds{1}}_{\rm cmp} = \hat{1}_N \oplus \hat{0}$ is the identity operation applied to the computational subspace of $N$ qubits, but without support outside the computational subspace ($\hat{0}$ is a zero matrix in the space outside the computational subspace), and therefore $\text{Tr}_{\rm cmp} [ \hat{L}^\dag(t) \mathds{1}_{\rm cmp} \hat{L}(t) ] \neq \text{Tr}_{\rm cmp} [ \hat{L}^\dag(t) \hat{L}(t) ]$. Note that the Hilbert-space dimension $d$ in \eqref{deltaF} is that of the computational subspace, i.e., $d=2^N$, no matter how many levels each qubit has beyond its computational subspace.

\paragraph*{Adiabatic time evolution.}

We now consider the consequences for the above results in the case where the Hamiltonian yields an adiabatic time evolution.
A state $\ket{\psi(t)}$ that evolves adiabatically under a time-dependent Hamiltonian $\hat{H}(t)$ has a time-evolution operator $\hat{U}(t) = {\cal T} \exp[-\frac{i}{\hbar}\int_{0}^{t} \hat H(t') \mathrm{d}t']$ that splits into a product~\cite{SuppMat}
\begin{equation}
   \hat{U}(t) = \hat{U}_\varphi (t) \hat{U}_\psi (t)
\end{equation}
of a phase operator $\hat{U}_\varphi (t)$ and a frame-transformation operator $\hat{U}_\psi (t)$. The phase operator $\hat{U}_\varphi (t) = \exp{i \hat{\varphi} (t)}$, with $\hat{\varphi} (t) = \hat{\theta} (t) + \hat{\gamma} (t)$, is defined by the dynamical and geometric (Berry) phase operators, $\hat{\theta} (t) = \sum_{m} \frac{-1}{\hbar} \int_{0}^{t} E_m(t') \mathrm{d}t' \ketbra{m(t)}{m(t)}$ and $\hat{\gamma} (t) = i \sum_{m} \int_{0}^{t} \braket{m(t')}{ \dot{m}(t')} \mathrm{d}t' \ketbra{m(t)}{m(t)}$. Here $\ket{m(t)}$ is the instantaneous eigenstate of $\hat{H}(t)$ with energy $E_m(t)$, and $\dot{m}(t)$ is the time derivative of the instantaneous eigenstate. The adiabatic frame-transformation operator, defined as
\begin{equation}
   \hat{U}_\psi (t) = \sum_{m} \ketbra{m(t)}{m(0)},
   \label{U_psi}
\end{equation}
is the operator that transfers the initial eigenstates to the instantaneous eigenstates at time $t$.

Remarkably, \eqref{deltaF} is independent of the phase operator $\hat{U}_\varphi (t)$ \emph{if the computational states are assumed to be initial eigenstates of $\hat H(t)$}. This is a consequence of the fact that the projected traces in \eqref{deltaF} are independent of the phase operator. To understand this, we note that for any operator $\hat{O}$ that adiabatically evolves as $\hat{O}(t) = \hat{U}^\dagger(t) \hat{O} \hat{U}(t)$, it holds that
\begin{equation}
   \Tr_{\rm cmp} [\hat{O}(t)] = \sum_{m \in \rm cmp} \braket{m(t)}{ \mathrm{e}^{-i \varphi_m(t) } \hat{O} \mathrm{e}^{i \varphi_m(t) } | m(t)},
\end{equation}
such that the scalar phase factors $\mathrm{e}^{\pm i \varphi_m(t) }$ cancel out~\cite{SuppMat}. Consequently, it is exact to simplify the time evolution in \eqref{deltaF} by replacing $\hat{L}(t) \to \hat{L}_\psi(t)$:
\begin{equation}
   \delta F[\hat{L}(t)] = \delta F[\hat{L}_\psi(t)],
   \label{deltaF_psi}
\end{equation}
where $\hat{L}_\psi(t) = \hat{U}_\psi^\dagger(t) \hat{L} \hat{U}_\psi(t)$ is the jump operator in the Heisenberg picture of the instantaneous eigenstates. 

\paragraph*{Perturbation theory for adiabatic state transfer.}
For general adiabatic state transfers, the frame-transformation operator $\hat{U}_\psi (t)$ is normally only known numerically. While $\hat{U}_\psi (t)$ already is significantly cheaper to compute, e.g., through exact diagonalization, compared to integrating the full master equation in \eqref{master}, we also insist on obtaining analytical formulae for the fidelity reduction from \eqref{deltaF_psi}. The analytical formulae help guide the intuition by revealing the dependence of the fidelity reduction on such parameters as coupling strengths $g$ and detunings $\Delta$, which are important in designing quantum gates.

To derive analytical formulae in terms of Hamiltonian parameters, we note that traces defining $\delta F[\hat{L}_\psi(t)]$ from \eqref{deltaF_psi} can be expressed in the bare eigenstates. These states are the eigenstates of the uncoupled Hamiltonian $\hat{H}_\mathrm{B}(t) \ket{m_\mathrm{B}(t)} = E_{\mathrm{B},m}(t) \ket{m_\mathrm{B}(t)}$, where $\hat H(t) = \hat H_\mathrm{B}(t) + \hat V(t)$, and $\hat V(t)$ denotes the couplings between the qubits. The key idea is that we can construct, similar to $\hat U_\psi(t)$, a frame-transformation operator
\begin{equation}
    \hat U_\mathrm{B}(t) = \sum_{m} \ketbra{m(t)}{m_\mathrm{B}(0)}
    \label{U_B}
\end{equation}
that transforms the initial bare eigenstates to the instantaneous eigenstates at time $t$. It fulfills the equation $ \hat U_\mathrm{B}(t) = \hat U_\psi(t) \hat U_\mathrm{B}(0)$, the inversion of which is used to find~\cite{SuppMat}
\begin{equation}
    \delta F[\hat{L}_\psi(t)] = \delta F_\mathrm{B}[\hat{L}_\mathrm{B}(t)],
   \label{deltaF_B}
\end{equation}
where the subscript in $\delta F_\mathrm{B}$ denotes that the trace is over the bare computational subspace, defined as $\Tr_{\rm B}[\cdot] = \sum_{m \in \mathrm{cmp}} \brakket{m_\mathrm{B}(0)}{ \cdot }{m_\mathrm{B}(0)}$, and $\hat{L}_\mathrm{B}(t) = \hat{U}_\mathrm{B}^\dagger(t) \hat{L} \hat{U}_\mathrm{B}(t)$. 

Using \eqref{deltaF_B}, we construct a perturbative expansion of the instantaneous eigenstates $\ket{m(t)}$ in terms of the initial bare eigenstates $\ket{m_\mathrm{B}(0)}$. The expansion is generated by a Schrieffer--Wolff (SW) transformation \cite{schriefferRelationAndersonKondo1966, bravyiSchriefferWolffTransformation2011}: $\ket{m(t)} = \exp{[-\hat{S}(t)]} \ket{m_\mathrm{B}(0)}$, where $\hat{S}(t)$, implicitly defined by
\begin{equation}
   \mathrm{e}^{\hat{S}(t)} \hat{H}(t) \mathrm{e}^{-\hat{S}(t)} = \hat{D}(t),
   \label{SW_equation}
\end{equation}
is the anti-hermitian generator that transforms $\hat{H}(t)$ into the diagonal Hamiltonian $\hat{D}(t)$. Note that diagonal, here, is with respect to the initial bare eigenstates $\ket{m_\mathrm{B}(0)}$. Perturbatively in $||{\hat{S}(t)}|| \ll 1$, \eqref{SW_equation} can be expanded with the Baker--Campbell--Hausdorff formula, giving the perturbative condition
\begin{equation}
   [\hat{S}(t), \hat{H}_\mathrm{B}(t)] = -\hat{V}(t).
   \label{SW_equation_perturbative}
\end{equation}

With the $\hat{S}(t)$ in hand, inserting the SW transformation into \eqref{U_B} reveals that $\hat{U}_\mathrm{B} (t) = \exp{[-\hat{S}(t)]}$, after the sum has collapsed using the resolution of the identity. Therefore, the jump operator in the Heisenberg picture of the instantaneous eigenstates has the perturbative expansion
\begin{equation}
   \hat{L}_\mathrm{B}(t) = \mathrm{e}^{\hat{S}(t)} \hat{L} \mathrm{e}^{-\hat{S}(t)} = \hat{L} + [\hat{S}(t), \hat{L}] + \mathcal{O}[\hat{S}^2(t)].
   \label{L_psi_perturbative}
\end{equation}

\paragraph*{Adiabatic CZ gates using a tunable coupler.}


For a concrete illustrative example, we consider the setup shown in~\figpanel{fig:gate_system}{a}, consisting of two fixed-frequency qubits coupled via a tunable coupler. The coupler frequency $\omega_c(t)$ is controlled by an external magnetic flux, enabling an adiabatic CZ gate by temporarily bringing the coupler close to resonance with the qubits, thereby accumulating a conditional phase. 
We model the system with the bosonic Hamiltonian ($\hbar = 1$)~\cite{Krantz2019}
\begin{align}
     \hat{H}(t) &= \sum_{i \in \{q_1,q_2,c\}} \mleft( \omega_i \hat{a}_i^\dagger \hat{a}_i + \frac{\alpha_i}{2} \hat{a}_i^\dagger \hat{a}_i^\dagger \hat{a}_i \hat{a}_i \mright) \notag \\
     &+ \sum_{i \neq j} g_{ij}(\hat{a}_i + \hat{a}_i^\dagger)(\hat{a}_j + \hat{a}_j^\dagger), \label{Hamiltonian-qubit-coupler-qubit}
\end{align}
with frequencies $\omega_i$, anharmonicities $\alpha_i$, and capacitive coupling strengths $g_{ij}$. 
The coupler frequency is $\omega_c(t) = \omega_{c,0}\sqrt{\left|\cos\!\left(\pi\Phi_{\mathrm{ext}}(t)\right)\right|}$, with $\omega_{c,0}$ the maximum coupler frequency and $\Phi_{\mathrm{ext}}(t)$
the applied flux (in units of the flux quantum). The qubit-coupler couplings inherit time dependency from $g_{ic} = g_{ic}[\sqrt{\omega_c(t)}]$~\cite{SuppMat}. 

\begin{figure*}
\includegraphics[width=\textwidth]{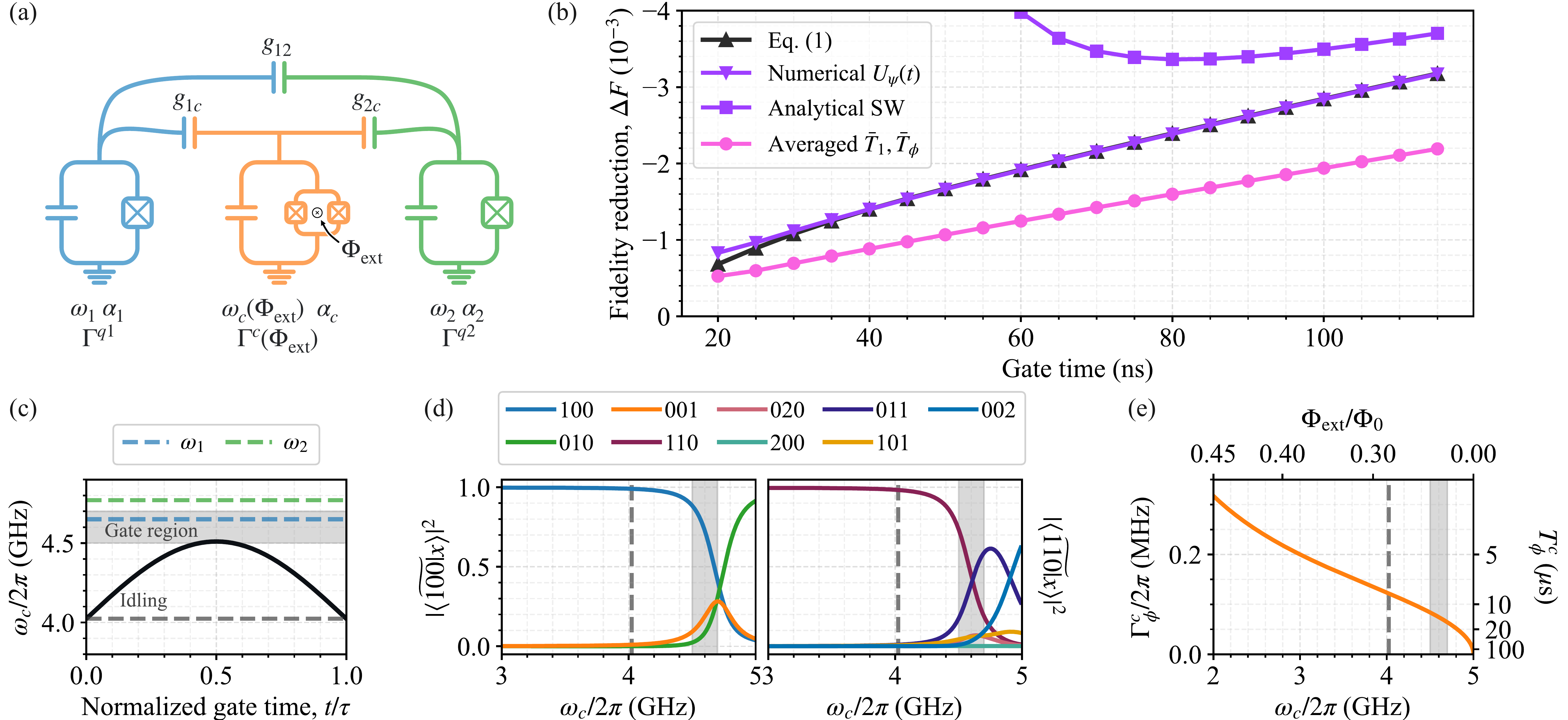}
  \caption{\textbf{Fidelity reduction for an adiabatic CZ gate in a two-qubit-coupler system.}
  \textbf{(a)} Circuit schematic showing two fixed-frequency qubits (blue, green) coupled via a tunable coupler (orange).
  \textbf{(b)} Fidelity reduction $ \Delta F = \bar{F} - 1$ as a function of gate time, evaluated numerically using \eqref{fidelity_integral} (black) and $U_\psi(t)$ (purple triangles), analytically using \eqref{eq:fidelity_reduction} from SW perturbation (purple squares), and with time-averaged decay rates (pink).
  \textbf{(c)} Time dependence of the coupler frequency during a sinusoidal flux pulse.
  \textbf{(d)} Hybridization of eigenstates, shown as the overlap between dressed and bare states as a function of coupler frequency.
  \textbf{(e)} Flux-noise sensitivity of the coupler, expressed as the dephasing rate $\Gamma_\phi^c$ and the corresponding coherence time $T_\phi^c$. Simulation parameters: $T_1^{q_i} = T_\phi^{q_i} = \SI{100}{\micro\second}$, $T_1^c = \SI{20}{\micro\second}$, and coupler dephasing giving $T_\phi^c = \SI{5}{\micro\second}$ at $\omega_c/2\pi = \SI{3}{\giga\hertz}$; further details in~\cite{SuppMat}.}
  \label{fig:gate_system}
\end{figure*}

In the dispersive regime $g_{ic}/\Delta_{ic}\ll1$, where $\Delta_{ic}=\omega_i-\omega_c$, ~\eqref{L_psi_perturbative} allows a perturbative evaluation of the instantaneous jump operators. For each mode, the decoherence processes are relaxation ($\hat{L}_1^i=\hat{a}_i$; rate $\Gamma_1^i$) and pure dephasing ($\hat{L}_\phi^i=\hat{a}_i^\dagger \hat{a}_i$; coherences decay at rate $2\Gamma_\phi^i$). These time-dependent operators are then inserted into the fidelity-reduction formula to obtain the integrand of~\eqref{fidelity}, i.e., the fidelity-reduction rate
\begin{align}
&\frac{\mathrm{d} \bar{F}}{\mathrm{d t}} = -\frac{2}{5}\sum_{i=1,2}\Gamma_{1}^{q_i} \mleft( 1 -\frac{g_{ic}^2}{\Delta_{ic}^2} \mright) -\frac{2}{5}\Gamma_{1}^{c} \mleft( \frac{g_{1c}^2}{\Delta_{1c}^2}+\frac{g_{2c}^2}{\Delta_{2c}^2} \mright) \nonumber \\
& -\frac{2}{5}\sum_{i=1,2}\Gamma_{\phi}^{q_i} \mleft( 1 + \frac{g_{ic}^2}{2 \Delta_{ic}^2} + \frac{5g_{12}^2}{\Delta _{12}^2} \mright) - \Gamma_{\phi}^{c} \mleft( \frac{g_{1c}^2}{\Delta_{1c}^2}+\frac{g_{2c}^2}{\Delta_{2c}^2} \mright).
\label{eq:fidelity_reduction}
\end{align}
For clarity, anharmonicities are neglected here; the full expression including their contribution is provided in~\cite{SuppMat}. The time dependence of the dissipation rates and detunings induced by flux tuning is also omitted.

This formula shows that the fidelity-reduction rate is given by a weighted sum of the different decay rates. The weights are set by the hybridization between the modes (which yields dressed eigenstates) and scale as $g_{ic}^2/\Delta_{ic}^2$ (assuming negligible $g_{12}$). In the limit $g_{ic}/\Delta_{ic} \to 0$, hybridization is suppressed, the dynamics remain confined to the computational subspace, and the result reduces to $-2\Gamma^{q_i}_{1,\phi}/5$, the known expression for two-qubit gates that do not leave this subspace~\cite{Abad2022}.
However, for finite hybridization, the eigenmodes become partially mixed [see~\figpanel{fig:gate_system}{d}], and each qubit inherits a contribution from the coupler decay. 

Equation~(\ref{eq:fidelity_reduction}) shows that the fidelity-reduction rate increases with qubit-coupler couplings, scaling approximately as $\mathrm{d} \bar{F} / \mathrm{d} t \sim g_{ic}^2 / \Delta_{ic}^2$. At the same time, increasing the coupling yields a faster gate, since the effective ZZ interaction scales as $\zeta \sim g_{ic}^4/\Delta_{ic}^3$~\cite{Sung2021, Chu2021, Fors2024}, leading to a gate time $\tau \sim 1/\zeta \sim \Delta_{ic}^3/g_{ic}^4$. The total fidelity reduction is obtained by integrating the instantaneous error over the gate duration [see~\eqref{fidelity}], yielding $\mathrm{d} \bar{F} / \mathrm{d} t \times \tau \sim \Delta_{ic}/g_{ic}^2$. Thus, although stronger coupling increases the fidelity-reduction rate, the final fidelity is increased because the gate time decreases more rapidly. In practice, however, the coupling strength cannot be increased arbitrarily, as it is limited by experimental constraints such as microwave crosstalk.

We can obtain more information about the fidelity reduction by building upon the framework introduced in Ref.~\cite{Abad2025} for leakage outside the computational subspace. Because adiabatic evolution preserves the instantaneous eigenbasis, we can isolate and evaluate the state-specific transition rates into distinct non-computational states driven by decoherence. Applying this formalism here, we derive explicit analytical expressions for state-specific and total leakage~\cite{SuppMat}, showing that the accumulated leakage is governed exclusively by the pure dephasing.
 
The perturbative theory used to obtain \eqref{eq:fidelity_reduction} is valid in the dispersive regime, where hybridization is weak. To treat regimes with stronger hybridization, we evaluate the fidelity reduction in \eqref{deltaF} numerically within the same framework by constructing the time-dependent operator $U_\psi(t)$ from the instantaneous eigenbasis of the Hamiltonian. We consider a flux pulse that produces a sinusoidal modulation of the coupler frequency [\figpanel{fig:gate_system}{c}], which sets the time dependence of the Hamiltonian in~\eqref{Hamiltonian-qubit-coupler-qubit}. Diagonalizing the Hamiltonian at each time yields $U_\psi(t)$, from which the time-dependent jump operators are obtained via \eqref{L-time-dependent} and inserted into \eqref{fidelity} to compute fidelity. To capture flux-dependent noise sensitivity, we use $\Gamma_\phi^c(t) \propto |\partial \omega_c / \partial \Phi|$, as in Ref.~\cite{Ithier2005}. The resulting fidelity reduction is shown as the purple line with triangle markers in \figpanel{fig:gate_system}{b}.

To validate this approach, we also compute the fidelity numerically using  \eqref{fidelity_integral} (black line with triangle markers). For this, we use the generalized formulation of Ref.~\cite{wood2018} (see~\cite{SuppMat}), which is not restricted to trace-preserving evolution within the computational subspace~\cite{Nielsen2002}. This is important here, since pure dephasing can lead to population loss from the computational subspace.
We also note that \eqref{fidelity_integral} includes all sources of fidelity reduction, both decoherence ($T_1$ and $T_\phi$) and coherent errors (e.g., leakage or imperfect gate operations). To extract the contribution from decoherence, we additionally compute the fidelity in the absence of losses. The difference between the fidelity with and without losses gives the fidelity reduction due to decoherence.

We find excellent agreement between the numerical evaluation based on \eqref{fidelity_integral} and the above approach, demonstrating that the $U_\psi(t)$ framework remains valid beyond the dispersive regime. Small deviations appear for short gate times ($\tau \lesssim \SI{25}{\nano\second}$), where the evolution becomes too rapid for the adiabatic approximation to hold.

The analytical SW result obtained from integrating~\eqref{eq:fidelity_reduction} is shown as the purple line with square markers in~\figpanel{fig:gate_system}{b}. It reproduces the overall behavior, but deviations become apparent as the system approaches avoided level crossings. In this regime, the assumption $g_{ic}/\Delta_{ic} \ll 1$ underlying the perturbative expansion is no longer valid, and the eigenmodes become strongly hybridized. 

We also evaluate the time-averaged error estimate introduced in Ref.~\cite{an2025}, $\Delta F = -\frac{2}{5} \sum_{i} (1/\bar{T}_1^{q_i} + 1/\bar{T}_{\varphi}^{q_i})\tau$, 
where the coherence times are averaged over the pulse. This expression corresponds to the standard result for two-qubit gates confined to the computational subspace~\cite{Abad2022}, but with the decay rates replaced by their averages along the trajectory.
This approach captures hybridization only through averaged decay rates and therefore significantly underestimates the accumulated error (pink line with circle markers).

\paragraph*{Conclusion and outlook.}

Time-dependent dissipation is an important feature of tunable quantum-computing architectures, where qubit or coupler frequencies are dynamically modulated, and in other quantum technologies. In this Letter, we extended the framework introduced in Refs.~\cite{Abad2022, Abad2025} to derive general analytical formulae quantifying the fidelity reduction of quantum operations in the presence of such time-dependent dissipation. We then applied this framework to adiabatic evolution by expressing the time-dependent jump operators in the instantaneous eigenbasis, which allows for a perturbative treatment using a Schrieffer--Wolff transformation.

As a concrete example, we derived an analytical expression for the fidelity reduction of widely used adiabatic CZ gates, which captures the interplay between time-dependent dissipation and mode hybridization. Our analysis clarified an important trade-off: while the fidelity-reduction rate increases with the qubit-coupler coupling, the total accumulated error still decreases since the gate time is reduced more rapidly. This provides a practical guideline for optimizing gate performance;
other recommendations include staying near flux ``sweet spots'' to suppress dephasing and avoiding regions of strong hybridization.
In addition to analytical results, our framework also enabled efficient numerical evaluation of the fidelity reduction that was significantly less demanding than integrating the full master equation. 

As a next step, our formulae could be applied to other concrete examples, including both other gates in quantum computers and protocols for quantum sensing and quantum communication. We also note that our analysis is based on a Markovian description of dissipation derived from the master-equation formalism, and as such does not capture non-Markovian noise sources, which may become more prominent when qubit quality increases~\cite{Burkard2009, Connolly2024, Gullans2024, Wei2024, Odeh2025, Zhuang2026}. Extending our framework to account for such non-Markovian effects remains an interesting direction for future work. More generally, our analysis can be extend to correlated dissipation~\cite{Chu2021, Li2024}. Even if the noise is initially uncorrelated, the instantaneous operators already mix the modes and generate effective correlations, and these contributions must be taken into account when assigning decay rates.

\paragraph*{Acknowledgements.}

We thank the experimental quantum-computing team at Chalmers University of Technology for helpful discussions. We acknowledge support from the Knut and Alice Wallenberg Foundation through the Wallenberg Centre for Quantum Technology (WACQT). AFK is also supported by the Swedish Foundation for Strategic Research (grant numbers FFL21-0279 and FUS21-0063) and the Horizon Europe programme HORIZON-CL4-2022-QUANTUM-01-SGA via the project 101113946 OpenSuperQPlus100.



\bibliography{ReferencesFidelity}

@misc{SuppMat, note = {{Supplementary Material.}} }

@article{Nielsen2002,
  title = {A simple formula for the average gate fidelity of a quantum dynamical operation},
  author = {Nielsen, Michael A.},
  year = {2002},
  journal = {Physics Letters A},
  volume = {303},
  pages = {249--252},
  doi = {10.1016/S0375-9601(02)01272-0}
}

@article{Abad2025,
  title = {Impact of decoherence on the fidelity of quantum gates leaving the computational subspace},
  author = {Abad, Tahereh and Schattner, Yoni and Frisk Kockum, Anton and Johansson, Göran},
  year = {2025},
  journal = {Quantum},
  volume = {9},
  pages = {1684},
  doi = {10.22331/q-2025-04-03-1684}
}

@article{Abad2022,
  title = {Universal fidelity reduction of quantum operations from weak dissipation},
  author = {Abad, Tahereh and Fern\'{a}ndez-Pend\'{a}s, Jorge and Frisk Kockum, Anton and Johansson, Göran},
  year = {2022},
  journal = {Physical Review Letters},
  volume = {129},
  pages = {150504},
  doi = {10.1103/PhysRevLett.129.150504}
}

@misc{Fors2024,
  title = {{Comprehensive explanation of ZZ coupling in superconducting qubits}},
  author = {Pettersson Fors, Simon and Fern\'{a}ndez-Pend\'{a}s, Jorge and Frisk Kockum, Anton},
  archivePrefix={arXiv},
  eprint  = {2408.15402},
  year = {2024},
}

@article{Acharya2025,
author = {Acharya, Rajeev and others},
doi = {10.1038/s41586-024-08449-y},
journal = {Nature},
pages = {920--926},
title = {{Quantum error correction below the surface code threshold}},
volume = {638},
year = {2025}
}

@misc{Ransford2025,
archivePrefix = {arXiv},
author = {Ransford, Anthony and others},
eprint = {2511.05465},
title = {{Helios: A 98-qubit trapped-ion quantum computer}},
year = {2025}
}

@misc{Strohm2024,
archivePrefix = {arXiv},
author = {Strohm, Thomas and Wintersperger, Karen and Dommert, Florian and Basilewitsch, Daniel and Reuber, Georg and Hoursanov, Andrey and Ehmer, Thomas and Vodola, Davide and Luber, Sebastian},
eprint = {2405.11450},
title = {{Ion-Based Quantum Computing Hardware: Performance and End-User Perspective}},
year = {2024}
}

@article{Wintersperger2023,
author = {Wintersperger, Karen and Dommert, Florian and Ehmer, Thomas and Hoursanov, Andrey and Klepsch, Johannes and Mauerer, Wolfgang and Reuber, Georg and Strohm, Thomas and Yin, Ming and Luber, Sebastian},
doi = {10.1140/epjqt/s40507-023-00190-1},
journal = {EPJ Quantum Technology},
pages = {32},
title = {{Neutral atom quantum computing hardware: performance and end-user perspective}},
volume = {10},
year = {2023}
}

@article{Bluvstein2026,
author = {Bluvstein, Dolev and others},
doi = {10.1038/s41586-025-09848-5},
journal = {Nature},
title = {{A fault-tolerant neutral-atom architecture for universal quantum computation}},
pages = {39--46},
volume = {649},
year = {2026}
}

@article{Burkard2023,
author = {Burkard, Guido and Ladd, Thaddeus D. and Pan, Andrew and Nichol, John M. and Petta, Jason R.},
doi = {10.1103/RevModPhys.95.025003},
journal = {Reviews of Modern Physics},
pages = {025003},
title = {{Semiconductor spin qubits}},
volume = {95},
year = {2023}
}

@misc{HRL2026,
archivePrefix = {arXiv},
author = {{Members of the HRL Quantum Team and Collaborators}},
eprint = {2604.16216},
title = {{A digitally controlled silicon quantum processing unit}},
year = {2026}
}

@article{Slussarenko2019,
author = {Slussarenko, Sergei and Pryde, Geoff J.},
doi = {10.1063/1.5115814},
journal = {Applied Physics Reviews},
pages = {041303},
title = {{Photonic quantum information processing: A concise review}},
volume = {6},
year = {2019}
}

@article{AghaeeRad2025,
author = {{Aghaee Rad}, H. and others},
doi = {10.1038/s41586-024-08406-9},
journal = {Nature},
pages = {912--919},
title = {{Scaling and networking a modular photonic quantum computer}},
volume = {638},
year = {2025}
}

@misc{Kinos2021,
archivePrefix = {arXiv},
author = {Kinos, Adam and others},
eprint = {2103.15743},
title = {{Roadmap for Rare-earth Quantum Computing}},
year = {2021}
}

@book{Nielsen2000,
author = {Nielsen, M. A. and Chuang, I. L.},
publisher = {Cambridge University Press},
title = {{Quantum Computation and Quantum Information}},
year = {2000}
}

@misc{Mohseni2025,
archivePrefix = {arXiv},
author = {Mohseni, Masoud and others},
eprint = {2411.10406},
title = {{How to Build a Quantum Supercomputer: Scaling from Hundreds to Millions of Qubits}},
year = {2025}
}

@article{Schlosshauer2019,
author = {Schlosshauer, Maximilian},
doi = {10.1016/j.physrep.2019.10.001},
journal = {Physics Reports},
pages = {1--57},
title = {{Quantum decoherence}},
volume = {831},
year = {2019}
}

@article{Terhal2015,
author = {Terhal, B. M.},
doi = {10.1103/RevModPhys.87.307},
journal = {Reviews of Modern Physics},
pages = {307},
title = {{Quantum error correction for quantum memories}},
volume = {87},
year = {2015}
}

@article{Girvin2023,
author = {Girvin, Steven M.},
doi = {10.21468/SciPostPhysLectNotes.70},
journal = {SciPost Physics Lecture Notes},
pages = {70},
title = {{Introduction to quantum error correction and fault tolerance}},
year = {2023}
}

@article{Degen2017,
author = {Degen, C. L. and Reinhard, F. and Cappellaro, P.},
doi = {10.1103/RevModPhys.89.035002},
journal = {Reviews of Modern Physics},
pages = {035002},
title = {{Quantum sensing}},
volume = {89},
year = {2017}
}

@article{Montenegro2025,
author = {Montenegro, Victor and Mukhopadhyay, Chiranjib and Yousefjani, Rozhin and Sarkar, Saubhik and Mishra, Utkarsh and Paris, Matteo G.A. and Bayat, Abolfazl},
doi = {10.1016/j.physrep.2025.05.005},
journal = {Physics Reports},
pages = {1--62},
title = {{Review: Quantum metrology and sensing with many-body systems}},
volume = {1134},
year = {2025}
}

@misc{Khan2025,
archivePrefix = {arXiv},
author = {Khan, Saeed A. and Prabhu, Sridhar and Wright, Logan G. and McMahon, Peter L.},
eprint = {2507.16918},
title = {{Quantum Computational-Sensing Advantage}},
year = {2025}
}

@article{Gisin2007,
author = {Gisin, Nicolas and Thew, Rob},
doi = {10.1038/nphoton.2007.22},
journal = {Nature Photonics},
pages = {165},
title = {{Quantum communication}},
volume = {1},
year = {2007}
}

@misc{Hajdusek2023,
archivePrefix = {arXiv},
author = {Hajdu{\v{s}}ek, Michal and {Van Meter}, Rodney},
eprint = {2311.02367},
title = {{Quantum Communications}},
year = {2023}
}

@article{Hyyppa2024,
author = {Hyypp{\"{a}}, Eric and others},
doi = {10.1103/PRXQuantum.5.030353},
journal = {PRX Quantum},
pages = {030353},
title = {{Reducing Leakage of Single-Qubit Gates for Superconducting Quantum Processors Using Analytical Control Pulse Envelopes}},
volume = {5},
year = {2024}
}

@article{Ding2023,
author = {Ding, Leon and others},
doi = {10.1103/PhysRevX.13.031035},
journal = {Physical Review X},
pages = {031035},
title = {{High-Fidelity, Frequency-Flexible Two-Qubit Fluxonium Gates with a Transmon Coupler}},
volume = {13},
year = {2023}
}

@article{Neeley2010,
author = {Neeley, M. and others},
doi = {10.1038/nature09418},
journal = {Nature},
pages = {570},
title = {{Generation of three-qubit entangled states using superconducting phase qubits}},
volume = {467},
year = {2010}
}

@article{Chen2014,
author = {Chen, Y. and others},
doi = {10.1103/PhysRevLett.113.220502},
journal = {Physical Review Letters},
pages = {220502},
title = {{Qubit Architecture with High Coherence and Fast Tunable Coupling}},
volume = {113},
year = {2014}
}

@article{Barends2019,
author = {Barends, R. and others},
doi = {10.1103/PhysRevLett.123.210501},
journal = {Physical Review Letters},
pages = {210501},
title = {{Diabatic Gates for Frequency-Tunable Superconducting Qubits}},
volume = {123},
year = {2019}
}

@article{Ding2025,
author = {Ding, Qi and Oppenheim, Alan V. and Boufounos, Petros T. and Gustavsson, Simon and Grover, Jeffrey A. and Baran, Thomas A. and Oliver, William D.},
doi = {10.1103/yskp-mfcr},
journal = {Physical Review Applied},
pages = {064013},
title = {{Pulse design of baseband flux control for adiabatic controlled-phase gates in superconducting circuits}},
volume = {23},
year = {2025}
}

@article{Haffner2008,
author = {H{\"{a}}ffner, H. and Roos, C. F. and Blatt, R.},
doi = {10.1016/j.physrep.2008.09.003},
journal = {Physics Reports},
pages = {155--203},
title = {{Quantum computing with trapped ions}},
volume = {469},
year = {2008}
}

@article{Chatterjee2021,
author = {Chatterjee, Anasua and Stevenson, Paul and {De Franceschi}, Silvano and Morello, Andrea and de Leon, Nathalie P. and Kuemmeth, Ferdinand},
doi = {10.1038/s42254-021-00283-9},
journal = {Nature Reviews Physics},
pages = {157--177},
title = {{Semiconductor qubits in practice}},
volume = {3},
year = {2021}
}

@article{Wendin2017,
author = {Wendin, G.},
doi = {10.1088/1361-6633/aa7e1a},
journal = {Reports on Progress in Physics},
pages = {106001},
title = {{Quantum information processing with superconducting circuits: a review}},
volume = {80},
year = {2017}
}

@article{Krantz2019,
author = {Krantz, P. and Kjaergaard, M. and Yan, F. and Orlando, T. P. and Gustavsson, S. and Oliver, W. D.},
doi = {10.1063/1.5089550},
journal = {Applied Physics Reviews},
pages = {021318},
title = {{A quantum engineer's guide to superconducting qubits}},
volume = {6},
year = {2019}
}

@article{Sung2021,
author = {Sung, Youngkyu and others},
doi = {10.1103/PhysRevX.11.021058},
journal = {Physical Review X},
pages = {021058},
title = {{Realization of High-Fidelity CZ and ZZ-Free iSWAP Gates with a Tunable Coupler}},
volume = {11},
year = {2021}
}

@article{Flamini2019,
author = {Flamini, F. and Spagnolo, N. and Sciarrino, F.},
doi = {10.1088/1361-6633/aad5b2},
journal = {Reports on Progress in Physics},
pages = {016001},
title = {{Photonic quantum information processing: a review}},
volume = {82},
year = {2019}
}

@article{Bruzewicz2019,
author = {Bruzewicz, Colin D. and Chiaverini, John and McConnell, Robert and Sage, Jeremy M.},
doi = {10.1063/1.5088164},
journal = {Applied Physics Reviews},
pages = {021314},
title = {{Trapped-ion quantum computing: Progress and challenges}},
volume = {6},
year = {2019}
}

@article{Caldwell2018,
author = {Caldwell, S. A. and others},
doi = {10.1103/PhysRevApplied.10.034050},
journal = {Physical Review Applied},
pages = {034050},
title = {{Parametrically Activated Entangling Gates Using Transmon Qubits}},
volume = {10},
year = {2018}
}

@article{Saffman2010,
author = {Saffman, M. and Walker, T. G. and M{\o}lmer, K.},
doi = {10.1103/RevModPhys.82.2313},
journal = {Reviews of Modern Physics},
pages = {2313--2363},
title = {{Quantum information with Rydberg atoms}},
url = {https://journals.aps.org/rmp/pdf/10.1103/RevModPhys.82.2313},
volume = {82},
year = {2010}
}

@article{Lindblad1976,
author = {Lindblad, G.},
doi = {10.1007/BF01608499},
journal = {Communications in Mathematical Physics},
pages = {119},
title = {{On the generators of quantum dynamical semigroups}},
volume = {48},
year = {1976}
}

@article{wood2018,
  title={Quantification and characterization of leakage errors},
  author={Wood, Christopher J and Gambetta, Jay M},
  journal={Physical Review A},
  volume={97},
  pages={032306},
  year={2018},
doi = {10.1103/PhysRevA.97.032306}
}

@article{Li2024,
  title = {{Realization of High-Fidelity CZ Gate Based on a Double-Transmon Coupler}},
  author = {Li, Rui and Kubo, Kentaro and Ho, Yinghao and Yan, Zhiguang and Nakamura, Yasunobu and Goto, Hayato},
  journal = {Physical Review X},
  volume = {14},
  pages = {041050},
  year = {2024},
  doi = {10.1103/PhysRevX.14.041050}
}

@misc{Marxer2025,
      title={{Above 99.9\% Fidelity Single-Qubit Gates, Two-Qubit Gates, and Readout in a Single Superconducting Quantum Device}}, 
      author={Fabian Marxer and others},
      year={2025},
      eprint={2508.16437},
      archivePrefix={arXiv}
}

@misc{Chen2026,
  title={{Unlocking a fast adiabatic CZ gate and exact residual ZZ cancellation between fixed-frequency transmons using a floating tunable coupler}},
  author={Angela Q. Chen and Xian Wu and Sarah Strong and Stefano Poletto},
  year={2026},
  eprint={2604.05048},
  archivePrefix={arXiv}
}

@book{sakuraiModernQuantumMechanics2020,
  title = {Modern {{Quantum Mechanics}}},
  author = {Sakurai, J. J. and Napolitano, Jim},
  year = {2020},
  month = sep,
  journal = {Higher Education from Cambridge University Press},
  publisher = {Cambridge University Press},
  doi = {10.1017/9781108587280},
  url = {https://www.cambridge.org/highereducation/books/modern-quantum-mechanics/DF43277E8AEDF83CC12EA62887C277DC},
  urldate = {2024-11-06},
  isbn = {9781108587280},
  langid = {english}
}

@article{schriefferRelationAndersonKondo1966,
  title = {Relation between the {{Anderson}} and {{Kondo Hamiltonians}}},
  author = {Schrieffer, J. R. and Wolff, P. A.},
  year = {1966},
  month = sep,
  journal = {Physical Review},
  volume = {149},
  number = {2},
  pages = {491--492},
  publisher = {American Physical Society},
  doi = {10.1103/PhysRev.149.491},
  url = {https://link.aps.org/doi/10.1103/PhysRev.149.491},
  urldate = {2024-08-11}
}

@article{bravyiSchriefferWolffTransformation2011,
  title = {Schrieffer--{{Wolff}} Transformation for Quantum Many-Body Systems},
  author = {Bravyi, Sergey and DiVincenzo, David P. and Loss, Daniel},
  year = {2011},
  month = oct,
  journal = {Annals of Physics},
  volume = {326},
  number = {10},
  pages = {2793--2826},
  issn = {0003-4916},
  doi = {10.1016/j.aop.2011.06.004},
  url = {https://www.sciencedirect.com/science/article/pii/S0003491611001059},
  urldate = {2024-08-11}
}

@article{Ithier2005,
  title = {Decoherence in a superconducting quantum bit circuit},
  author = {Ithier, G. and Collin, E. and Joyez, P. and Meeson, P. J. and Vion, D. and Esteve, D. and Chiarello, F. and Shnirman, A. and Makhlin, Y. and Schriefl, J. and Sch\"on, G.},
  journal = {Physical Review B},
  volume = {72},
  pages = {134519},
  year = {2005},
  doi = {10.1103/PhysRevB.72.134519}
}

@article{Chu2021,
  title = {{Coupler-Assisted Controlled-Phase Gate with Enhanced Adiabaticity}},
  author = {Chu, Ji and Yan, Fei},
  journal = {Physical Review Applied},
  volume = {16},
  pages = {054020},
  year = {2021},
  doi = {10.1103/PhysRevApplied.16.054020}
}

@misc{an2025,
      title={{ZZ-Free Two-Transmon CZ Gate Mediated by a Fluxonium Coupler}}, 
      author={Junyoung An and Helin Zhang and Qi Ding and Leon Ding and Youngkyu Sung and Roni Winik and Junghyun Kim and Ilan T. Rosen and Kate Azar and Renee DePencier Piñero and Jeffrey M. Gertler and Michael Gingras and Bethany M. Niedzielski and Hannah Stickler and Mollie E. Schwartz and Joel {\^I}-j. Wang and Terry P. Orlando and Simon Gustavsson and Max Hays and Jeffrey A. Grover and Kyle Serniak and William D. Oliver},
      year={2025},
      eprint={2511.02115},
      archivePrefix={arXiv}
}

@article{Burkard2009,
author = {Burkard, Guido},
doi = {10.1103/PhysRevB.79.125317},
journal = {Physical Review B},
pages = {125317},
title = {{Non-Markovian qubit dynamics in the presence of 1/f noise}},
volume = {79},
year = {2009}
}

@article{Connolly2024,
author = {Connolly, Thomas and Kurilovich, Pavel D. and Diamond, Spencer and Nho, Heekun and B{\o}ttcher, Charlotte G. L. and Glazman, Leonid I. and Fatemi, Valla and Devoret, Michel H.},
doi = {10.1103/PhysRevLett.132.217001},
journal = {Physical Review Letters},
pages = {217001},
title = {{Coexistence of Nonequilibrium Density and Equilibrium Energy Distribution of Quasiparticles in a Superconducting Qubit}},
volume = {132},
year = {2024}
}

@article{Gullans2024,
author = {Gullans, M.J. and Caranti, M. and Mills, A.R. and Petta, J.R.},
doi = {10.1103/PRXQuantum.5.010306},
journal = {PRX Quantum},
pages = {010306},
title = {{Compressed Gate Characterization for Quantum Devices with Time-Correlated Noise}},
volume = {5},
year = {2024}
}

@article{Wei2024,
author = {Wei, Ken Xuan and Pritchett, Emily and Zajac, David M. and McKay, David C. and Merkel, Seth},
doi = {10.1103/PhysRevApplied.21.024018},
journal = {Physical Review Applied},
pages = {024018},
title = {{Characterizing non-Markovian off-resonant errors in quantum gates}},
volume = {21},
year = {2024}
}

@article{Odeh2025,
author = {Odeh, Mutasem and Godeneli, Kadircan and Li, Eric and Tangirala, Rohin and Zhou, Haoxin and Zhang, Xueyue and Zhang, Zi-Huai and Sipahigil, Alp},
doi = {10.1038/s41567-024-02740-5},
journal = {Nature Physics},
pages = {406},
title = {{Non-Markovian dynamics of a superconducting qubit in a phononic bandgap}},
volume = {21},
year = {2025}
}

@article{Zhuang2026,
author = {Zhuang, Ze-Tong and Rosenstock, Dario and Liu, Bao-Jie and Somoroff, Aaron and Manucharyan, Vladimir E. and Wang, Chen},
doi = {10.1038/s41467-026-69910-2},
journal = {Nature Communications},
pages = {3209},
title = {{Non-Markovian relaxation spectroscopy of fluxonium qubits}},
volume = {17},
year = {2026}
}

\clearpage
\onecolumngrid
\setcounter{section}{0}
\renewcommand{\thesection}{S\arabic{section}}
\setcounter{equation}{0}
\renewcommand{\theequation}{S\arabic{equation}}
\renewcommand{\thetable}{S\arabic{table}}
\renewcommand{\bibnumfmt}[1]{[S#1]}
\setcounter{page}{1}

%


\section{Derivation of average gate fidelity with time-dependent dissipation}
\label{sc:dyson_expansion_master_equation}

Here we present the details of the derivation leading to Eq.~(3) in the main text for a system with a time-dependent Hamiltonian and time-dependent dissipation. We start from the Lindblad master equation, Eq.~(2) in the main text, setting $\hbar = 1$ throughout this Supplementary Material:
\begin{equation}
    \dot{\rho}(t) = \mleft(\mathcal{S}[\hat{H}(t)] + \sum_{k=1}^{N_L} {\Gamma_k(t)} \mathcal{D}[\hat{L}_k]\mright)\rho(t),
    \label{eq:master_equation}
\end{equation}
where we define the superoperator $\mathcal{S}[\hat{H}] \rho = -i[\hat{H}, \rho]$ and assume that the time dependence of the dissipation enters through the rates $\Gamma_k(t)$. To describe the evolution in a compact way, we introduce the time-evolution superoperator $\mathcal{U}_L(t_2, t_1)$, defined by
\begin{equation}
   \rho(t_2) := \mathcal{U}_L(t_2, t_1)\rho(t_1),
   \label{eq:time_evolution_superoperator}
\end{equation}
which propagates the state from time $t_1$ to $t_2$. From the master equation, this superoperator obeys
\begin{equation}
   \dot{\mathcal{U}}_L(t, 0) = \mleft(\mathcal{S}[\hat{H}(t)] + \sum_{k=1}^{N_L} {\Gamma_k(t)} \mathcal{D}[\hat{L}_k]\mright)\mathcal{U}_L(t, 0).
\end{equation}

To proceed, we expand the time-evolution superoperator $\mathcal{U}_L(t, 0)$ perturbatively in the dissipation using a Dyson series. As a starting point, we consider the ideal case without dissipation, for which the evolution is governed solely by the Hamiltonian
\begin{equation}
   \dot{\mathcal{U}}^{(0)}_L(t, 0) = \mathcal{S}[\hat{H}(t)] \mathcal{U}^{(0)}_L(t, 0),
\end{equation}
where the index $(0)$ indicates zeroth order in the dissipation. Since we are interested in quantum operations, we assume that this unitary evolution $\mathcal{U}^{(0)}_L(t, 0)$ is known.

We then use this evolution to move to the interaction (Heisenberg) picture, defining
\begin{equation}
   \rho_I(t) = \mathcal{U}^{(0) \dag}_L(t, 0) \rho(t),
\end{equation}
which removes the purely unitary dynamics generated by the Hamiltonian. Inserting this into \eqref{eq:master_equation}, we obtain the Lindblad master equation in the interaction picture:
\begin{equation}
   \dot{\rho}_I(t) = \mleft( \sum_{k=1}^{N_L} {\Gamma_k(t)} \mathcal{D}_I[\hat{L}_k; t] \mright) \rho_I(t),\label{eq:master_equation_interaction_picture}
\end{equation}
where the dissipator in the interaction picture is given by
\begin{equation}
   \mathcal{D}_I[\hat{L}_k; t] = \mathcal{U}^{(0) \dag}_L(t, 0) \mathcal{D}[\hat{L}_k] \mathcal{U}^{(0)}_L(t, 0).
   \label{eq:dissipator_interaction_picture}
\end{equation}

The solution to \eqref{eq:master_equation_interaction_picture} can be written in terms of a time-evolution superoperator in the interaction picture as
\begin{equation}
    \rho_I(t_2) = \mathcal{U}_I(t_2, t_1) \rho_I(t_1),
\end{equation}
where the time-evolution superoperator in the interaction picture is formally the time-ordered exponential
\begin{equation}
  \mathcal{U}_I(t_2, t_1) = \mathcal{T} \exp \mleft[ \int_{t_1}^{t_2} \mathrm{d} t' \mleft( \sum_{k=1}^{N_L} {\Gamma_k(t')} \mathcal{D}_I[\hat{L}_k; t'] \mright) \mright],
\end{equation}
where $\mathcal{T}$ denotes time-ordering.

Transforming back to the Schr\"odinger picture, the solution to \eqref{eq:time_evolution_superoperator} can be written as
\begin{equation}
  \rho(t) = \mathcal{U}^{(0)}_L(t, 0) \mathcal{U}_I(t, 0) \rho(0),
\end{equation}
where $\mathcal{U}_L(t, 0) = \mathcal{U}^{(0)}_L(t, 0) \mathcal{U}_I(t, 0)$.

The time-ordered exponential can be expanded as a Dyson-series expansion, which provides a perturbative expansion in the dissipation:
\begin{equation}
  \mathcal{U}_I(t_2, t_1) = \sum_{n=0}^{\infty} \frac{1}{n!} \int_{t_1}^{t_2} \mathrm{d} t'_1 \cdots \int_{t_1}^{t_2} \mathrm{d} t'_n \mathcal{T}
  \mleft( \sum_{k=1}^{N_L} {\Gamma_k(t'_1)} \mathcal{D}_I[\hat{L}_k; t'_1] \mright)
  \cdots
  \mleft( \sum_{k=1}^{N_L} {\Gamma_k(t'_n)} \mathcal{D}_I[\hat{L}_k; t'_n] \mright),
\end{equation}
Keeping terms up to the first order in the dissipation, we obtain
\begin{equation}\label{eq:1st o expansion}
  \mathcal{U}_L(t, 0) = \mathcal{U}^{(0)}_L(t, 0) + \sum_{k=1}^{N_L} \int_{0}^{t} \mathrm{d} t' {\Gamma_k(t')} \mathcal{U}^{(0)}_L(t, t') \mathcal{D}[\hat{L}_k] \mathcal{U}^{(0)}_L(t', 0) + \mathcal{O}(\bar{\Gamma}_k^2 \tau^2),
\end{equation}
where we used the time-reversal and composition properties of the unitary evolution, $\mathcal{U}^{(0)}_L(t, 0) \mathcal{U}^{(0) \dag}_L(t', 0) = \mathcal{U}^{(0)}_L(t, 0) \mathcal{U}^{(0)}_L(0, t') = \mathcal{U}^{(0)}_L(t, t')$. Here, $\bar{\Gamma}_{k} = \frac{1}{\tau} \int_0^\tau dt \, \Gamma_k(t)$ denotes the  time-averaged rate for the $k$th dissipative process.

Following the same perturbative approach as in Ref.~\cite{Abad2022}, we directly obtain the average gate fidelity
\be \label{}
\bar{F} = 1 + \sum_{k = 1}^{N_L} \int_0^\tau \mathrm{d} t \, \Gamma_k(t) \delta F[\hat{L}_k (t)] + \mathcal{O}\mleft(\bar{\Gamma}_k^2 \tau^2 \mright),
\ee
where the only difference compared to Ref.~\cite{Abad2022} is that the dissipation rates $\Gamma_k(t)$ now enter as time-dependent kernels in the integral, rather than as constant prefactors. This corresponds to Eq.~(3) in the main text, where the fidelity-reduction integrand $\delta F[\hat{L}_k (t)]$ is given by Eq.~(5) in the main text, as shown in Ref.~\cite{Abad2025}.

\section{Fidelity reduction for adiabatic time evolution}
\label{sc:fidelity_reduction_adiabatic_time_evolution}
In this section, we provide the derivation of the fidelity reduction for adiabatic time evolution given in Eq.~(9) in the main text. The main result is that the fidelity reduction is independent of the phase operator $\hat{U}_\varphi (t)$, provided that the computational basis coincides  with the eigenstates of $ \hat H(0)$.

The phase operator is an integral part of Eq.~(6) in the main text, which is derived from the adiabatic theorem \cite{sakuraiModernQuantumMechanics2020}. This theorem applies when the Hamiltonian varies slowly compared to the energy gaps in the energy spectrum. In this regime, the system remains in its instantaneous eigenstate up to a phase. The dynamics are governed by the time-dependent Schr\"odinger equation
\begin{equation}
   \hat H(t) \ket{m(t)} = E_m(t) \ket{m(t)},
\end{equation}
where $\ket{m(t)}$ is the instantaneous eigenstate of the Hamiltonian $\hat H(t)$ with eigenvalue $E_m(t)$. The instantaneous eigenstates form an orthonormal basis and can therefore be used to expand any state as
\begin{equation}
    \ket{\psi(t)} = \sum_{m} c_m(t) \ket{m(t)} ,
    \label{eq:adiabatic_state}
\end{equation}
where the amplitudes are given by $c_m(t) = \braket{m(t)}{\psi(t)}$. Under adiabatic evolution, the amplitudes acquire only phase factors and evolve as
\begin{equation}
  c_m(t) = c_m(0) \e^{\i \theta_m(t)} \e^{\i \gamma_m(t)},
  \label{eq:adiabatic_theorem}
\end{equation}
where $\theta_m(t) = - \int_0^t E_m(t') \d t'$ is the dynamical phase and $\gamma_m(t) = \i \int_0^t \braket{m(t')}{\dot{m}(t')} \d t'$ is the geometric (Berry) phase. Here, $\ket{\dot{m}(t)} = \partial_t \ket{m(t)}$ denotes the time derivative of the instantaneous eigenstate.

The adiabatic approximation holds when transitions between different instantaneous eigenstates are suppressed. This is ensured by the adiabatic condition
\begin{equation}
  \frac{1}{\abs{ \i E_m(t) +  \braket{m(t)}{\dot{m}(t)}}} \abs{ \sum_{n \neq m} \frac{\brakket{m(t)}{\partial_t H}{n(t)}}{E_m(t) - E_n(t) } \frac{c_n(t)}{{\abs{c_m(t)}}} } \ll 1,
  \label{eq:adiabatic_condition}
\end{equation}
which quantifies the suppression of nonadiabatic couplings.

To derive Eq.~(6) in the main text, we start by noting that $c_m(0) \ket{m(t)} = \mleft( \ketbra{m(t)}{m(0)} \mright) \ket{\psi(0)}$. We then use the orthogonality of the instantaneous eigenstates to construct the dynamical and geometric phase operators
\begin{align}
  \hat{\theta} (t) &= \sum_{m} \theta_m \ketbra{m(t)}{m(t)} , \\
  \hat{\gamma} (t) &= \sum_{m} \gamma_m \ketbra{m(t)}{m(t)} ,
\end{align}
which are diagonal in the instantaneous eigenstates. In particular, they satisfy $\hat{\theta} (t) \ket{m(t)} = \theta_m(t) \ket{m(t)}$ and $\hat{\gamma} (t) \ket{m(t)} = \gamma_m(t) \ket{m(t)}$, with the dynamical and geometric phase, respectively, as eigenvalues.

Using these operators, \eqref{eq:adiabatic_state} can be written as
\begin{equation}
  \ket{\psi(t)} = \e^{\i[\hat \theta(t) + \hat \gamma(t)]} \mleft( \sum_m \ketbra{m(t)}{m(0)} \mright) \ket{\psi(0)}.
\end{equation}
We thus identify the phase operator $\hat{U}_\varphi (t) = \exp{\mathrm{i} \hat{\varphi} (t)}$, with $\hat{\varphi} (t) = \hat{\theta} (t) + \hat{\gamma} (t)$, and the frame-transformation operator $\hat{U}_\psi (t) = \sum_{m} \ketbra{m(t)}{m(0)}$. The time-evolution operator therefore factorizes as
\begin{equation}
  \hat{U}(t) = \hat{U}_\varphi(t) \hat{U}_\psi(t),
\end{equation}
such that $\ket{\psi(t)} = \hat{U}(t) \ket{\psi(0)}$. This corresponds to Eq.~(6) in the main text.

We now compute the fidelity reduction for adiabatic time evolution in Eq.~(9) in the main text. We assume that the the computational states coincide with the eigenstates of $\hat H(0)$, such that the projector onto the computational subspace is given by
\begin{equation} \label{I_cmp}
    \hat{\mathds{1}}_{\rm cmp} = \sum_{m \in \mathrm{cmp}} \ketbra{m(0)}{m(0)},
\end{equation}
where ``cmp'' denotes the set of states spanning the computational subspace.

The trace terms in Eq.~(5) in the main text are then invariant under the phase operator. For example,
\begin{equation}
\begin{aligned}
  \Tr_{\rm cmp}{\mleft[ \hat{L}(t) \mright]} 
  &= \sum_{m \in \mathrm{cmp}} \brakketsmall{m(0)}{ \hat{U}_\psi^\dagger (t) \hat{U}_\varphi^\dagger (t) \hat L \hat{U}_\varphi(t) \hat{U}_\psi(t) }{m(0)} \\
  &= \sum_{m \in \mathrm{cmp}} \brakketsmall{m(t)}{ \e^{-\i \varphi_m(t)} \hat L \e^{\i \varphi_m(t)} }{m(t)} \\
  &= \sum_{m \in \mathrm{cmp}} \brakketsmall{m(0)}{ \hat L_\psi(t) }{m(0)} = \Tr_{\rm cmp}{\mleft[ \hat{L}_\psi(t) \mright]},
\end{aligned}
\end{equation}
where the scalar phase factors $\e^{\pm \i \varphi_m(t)}$ cancel in the second line and $\hat{L}_\psi(t) = \hat{U}_\psi^\dagger(t) \hat{L} \hat{U}_\psi(t)$ denotes the jump operator in the Heisenberg picture of the instantaneous eigenstates.

The same argument applies to the remaining trace terms, yielding
\begin{equation}
   \Tr_{\rm cmp}{\mleft[ \hat{L}^\dag(t) \mright]} = \Tr_{\rm cmp}{\mleft[ \hat{L}_\psi^\dag(t) \mright]}
\end{equation}
and
\begin{equation}
    \Tr_{\rm cmp}{\mleft[ \hat{L}^\dag(t) \hat{L}(t) \mright]} = \Tr_{\rm cmp}{\mleft[ \hat{L}_\psi^\dag(t) \hat{L}_\psi(t) \mright]}.
\end{equation}

For the second term in Eq.~(5) in the main text, which includes the computational projector, a similar calculation gives
\begin{equation}
\begin{aligned}
    \text{Tr}_{\rm cmp} \mleft[ \hat{L}^\dag(t) \hat{\mathds{1}}_{\rm cmp} \hat{L}(t) \hat{\mathds{1}}_{\rm cmp} \mright]
    &= 
    \sum_{m \in \mathrm{cmp}} 
    \brakketsmall{m(0)}
    { 
      \hat{U}_\psi^\dagger (t) \hat{U}_\varphi^\dagger (t) \hat L^\dag \hat{U}_\varphi(t) \hat{U}_\psi(t)
      \hat{\mathds{1}}_{\rm cmp}
      \hat{U}_\psi^\dagger (t) \hat{U}_\varphi^\dagger (t) \hat L \hat{U}_\varphi(t) \hat{U}_\psi(t)
    }{m(0)} \\
    &=
    \sum_{m \in \mathrm{cmp}} 
    \brakketsmall{m(t)}
    { 
      \e^{-\i \varphi_m(t)}
      \hat L^\dag 
      \mleft(
      \sum_{n \in \mathrm{cmp}} \e^{-\i \varphi_n(t)} \ketbra{n(t)}{n(t)} \e^{\i \varphi_m(t)}
      \mright)
      \hat L
      \e^{\i \varphi_m(t)}
    }{m(t)} \\
    &=
    \sum_{m \in \mathrm{cmp}} 
    \brakketsmall{m(0)}
    { 
      \hat L^\dag_\psi(t)
      \hat{\mathds{1}}_{\rm cmp}
      \hat L_\psi(t)
      \hat{\mathds{1}}_{\rm cmp}
    }{m(0)} \\
    &=
    \text{Tr}_{\rm cmp} \mleft[ \hat{L}_\psi^\dag(t) \hat{\mathds{1}}_{\rm cmp} \hat{L}_\psi(t) \hat{\mathds{1}}_{\rm cmp} \mright],
\end{aligned}
\end{equation}
where the scalar phase factors $\e^{\pm \i \varphi_m(t)}$ and $\e^{\pm \i \varphi_n(t)}$ cancel in the second line. Inserting these results into Eq.~(5) in the main text, we obtain
\begin{equation}
   \delta F[\hat{L}(t)] = \delta F[\hat{L}_\psi(t)],
   \label{}
\end{equation}
which corresponds to Eq.~(9) in the main text.

\section{Fidelity reduction in the bare eigenstate representation}
\label{sc:fidelity_reduction_bare_basis}
Here we derive Eq.~(11) in the main text, where the fidelity reduction $\delta F[\hat{L}_\psi(t)]$ is expressed in terms of traces over the bare computational subspace, denoted $\delta F_\mathrm{B}[\hat{L}_B(t)]$. We provide the intermediate steps of this rewriting and give the explicit definition of $F_\mathrm{B}$.

Starting from the frame-transformation operator $\hat U_\mathrm{B}(t)$ in Eq.~(12) in the main text, which maps the bare eigenstates to the instantaneous eigenstates, we first write
\begin{equation}
    \hat U_\psi(t) \hat U_\mathrm{B}(0) = \sum_{m,m'} \ket{m'(t)} \braket{m'(0)}{m(0)} \bra{m_\mathrm{B}(0)} = \hat U_\mathrm{B}(t),
\end{equation}
which follows from the orthonormality relation $\delta_{mm'} = \braket{m'(0)}{m(0)}$.

This identity simply reflects that transforming from the bare eigenstates to the initial instantaneous eigenstates and subsequently to the instantaneous eigenstates at time $t$, is equivalent to directly transforming from the bare eigenstates to the instantaneous eigenstates at time $t$. It thus follows that
\begin{equation}
    \hat U_\psi(t) = \hat U_\mathrm{B}(t) \hat U_\mathrm{B}^\dag(0).
\end{equation}

We substitute $U_\psi(t)$ in $\delta F_N[\hat{L}_\psi(t)]$ and find
\begin{equation}
\begin{aligned}
    \Tr_{\rm cmp}{\mleft[ \hat{L}_\psi(t) \mright]} &=
    \sum_{m \in \mathrm{cmp}} \brakketsmall{m(0)}{ \hat U_\mathrm{B}(0) \hat U_\mathrm{B}^\dag(t) \hat L \hat U_\mathrm{B}(t) \hat U_\mathrm{B}^\dag(0) }{m(0)} \\
    &= \sum_{m \in \mathrm{cmp}} \brakketsmall{m_\mathrm{B}(0)}{ \hat L_\mathrm{B}(t) }{m_\mathrm{B}(0)}
    = \Tr_{\rm B} {\mleft[ \hat{L}_\mathrm{B}(t) \mright]},
\end{aligned}
\end{equation}
where $\hat{L}_\mathrm{B}(t) = \hat{U}_\mathrm{B}^\dagger(t) \hat{L} \hat{U}_\mathrm{B}(t)$ and $\Tr_{\rm B}$ denotes the trace over the bare computational subspace.

The same argument directly yields
\begin{equation}
   \Tr_{\rm cmp}{\mleft[ \hat{L}^\dag(t) \mright]} = \Tr_{\rm B}{\mleft[ \hat{L}_\mathrm{B}^\dag(t) \mright]}, 
\end{equation}
and
\begin{equation}
  \Tr_{\rm cmp}{\mleft[ \hat{L}^\dag(t) \hat{L}(t) \mright]} = \Tr_{\rm B}{\mleft[ \hat{L}_\mathrm{B}^\dag(t) \hat{L}_\mathrm{B}(t) \mright]}.
\end{equation}
A similar calculation, using \eqref{I_cmp}, yields
\begin{equation}
\begin{aligned}
    \text{Tr}_{\rm cmp} \mleft[ \hat{L}_\psi^\dag(t) \hat{\mathds{1}}_{\rm cmp} \hat{L}_\psi(t) \hat{\mathds{1}}_{\rm cmp} \mright] 
    &= 
    \sum_{m \in \mathrm{cmp}} 
    \brakketsmall{m(0)}
    { 
      \hat U_\mathrm{B}(0) \hat U_\mathrm{B}^\dag(t)
      \hat L^\dag
      \hat U_\mathrm{B}(t) \hat U_\mathrm{B}^\dag(0)
      \hat{\mathds{1}}_{\rm cmp}
      \hat U_\mathrm{B}(0) \hat U_\mathrm{B}^\dag(t)
      \hat L
      \hat U_\mathrm{B}(t) \hat U_\mathrm{B}^\dag(0)
    }{m(0)} \\
    &= 
    \sum_{m \in \mathrm{cmp}} 
    \brakketsmall{m_\mathrm{B}(0)}
    { 
      \hat L_\mathrm{B}^\dag(t)
      \hat{\mathds{1}}_{\rm B}
      \hat L_\mathrm{B}(t)
      \hat{\mathds{1}}_{\rm B}
    }{m_\mathrm{B}(0)} =
    \text{Tr}_{\rm B} \mleft[ \hat{L}_\mathrm{B}^\dag(t) \hat{\mathds{1}}_{\rm B} \hat{L}_\mathrm{B} \hat{\mathds{1}}_{\rm B} \mright], 
\end{aligned}
\end{equation}
where $\hat{\mathds{1}}_{\rm B} = \sum_{m \in \mathrm{cmp}} \ketbra{m_\mathrm{B}(0)}{m_\mathrm{B}(0)}$ is the projector onto the bare computational subspace.

Inserting these expressions into Eq.~(5) in the main text, we obtain
\begin{equation}
\begin{aligned}
    \delta F[\hat{L}_\psi(t)] &= \frac{1}{d(d+1)} \Tr_{\rm B}{\mleft[ \hat{L}_\mathrm{B}^\dag(t) \mright]} \Tr_{\rm B}{\mleft[ \hat{L}_\mathrm{B}(t) \mright]}
    + \frac{1}{d(d+1)} \text{Tr}_{\rm B} \mleft[ \hat{L}_\mathrm{B}^\dag(t) \hat{\mathds{1}}_{\rm B} \hat{L}_\mathrm{B}(t) \hat{\mathds{1}}_{\rm B} \mright] \\
    &- \frac{1}{d} \Tr_{\rm B}{\mleft[ \hat{L}_\mathrm{B}^\dag(t) \, \hat{L}_\mathrm{B}(t) \mright]} = \delta F_\mathrm{B}[\hat{L}_B(t)].
\end{aligned}
\end{equation}
This corresponds to Eq.~(11) in the main text.

\section{Fidelity reduction of adiabatic CZ gates}
Here we present the details of the derivation leading to Eq.~(16) in the main text. To evaluate the fidelity reduction, we express the jump operators in the Heisenberg picture,
\begin{equation}
    \hat{L}(t) = \hat{U}^\dagger(t)\hat{L}\hat{U}(t),
\end{equation}
where $\hat{U}(t)$ is the time-evolution operator. For adiabatic state transfer, we use a SW transformation to obtain this evolution, writing $\hat{U}(t) = e^{-\hat{S}}$. The anti-Hermitian generator is given by~\cite{Fors2024}
\begin{equation} \label{eq:SW_generator}
\hat{S} = \sum_{ij} g_{ij}\,\hat{a}_i^\dagger \frac{1}{\hat{D}_{ij}} \hat{a}_j.
\end{equation}
Here, $g_{ij}$ are the coupling strengths, $\hat{a}_i^\dagger$ ($\hat{a}_i$) are the standard bosonic creation (annihilation) operators, and the operator-valued denominator is
\begin{equation}
    \hat{D}_{ij} = \Delta_{ij} + \alpha_i \hat{a}_i^\dagger \hat{a}_i - \alpha_j \hat{a}_j^\dagger \hat{a}_j,
\end{equation}
with detuning $\Delta_{ij} = \omega_i - \omega_j$. 

In the weakly nonlinear regime, where the anharmonicities satisfy $|\alpha_i|, |\alpha_j| \ll |\Delta_{ij}|$, we approximate $D_{ij} \simeq \Delta_{ij}$ to obtain compact analytical expressions. We then expand the annihilation operator in the dressed frame to second order in $\hat{S}$ using the Baker--Campbell--Hausdorff (BCH) lemma,
\begin{equation} \label{eq:ak_expansion}
\hat{a}_k(t) = e^{\hat{S}} \hat{a}_k e^{-\hat{S}} = \hat{a}_k + [\hat{S}, \hat{a}_k] + \frac{1}{2}[\hat{S}, [\hat{S}, \hat{a}_k]] + \mathcal{O}(\hat{S}^3).
\end{equation}

Using the bosonic commutation relations $[\hat{a}_j, \hat{a}_k] = 0$ and $[\hat{a}_i, \hat{a}_k^\dagger] = \delta_{ik}$, the first-order term becomes
\begin{equation} \label{eq:ak_first_order}
\hat{a}_k^{(1)} \equiv [\hat{S}, \hat{a}_k] = \sum_{ij} \frac{g_{ij}}{\Delta_{ij}} [\hat{a}_i^\dagger \hat{a}_j, \hat{a}_k] = - \sum_j \frac{g_{kj}}{\Delta_{kj}} \hat{a}_j,
\end{equation}
and the second-order contribution reads
\begin{equation} \label{eq:ak_second_order}
\hat{a}_k^{(2)} \equiv \frac{1}{2} [\hat{S}, [\hat{S}, \hat{a}_k]] = \frac{1}{2} \sum_{i,j} \frac{g_{ki}}{\Delta_{ki}} \frac{g_{ij}}{\Delta_{ij}} \hat{a}_j.
\end{equation}
The dressed operator to second order is thus given by $\hat{a}_k(t) = \hat{a}_k + \hat{a}_k^{(1)} + \hat{a}_k^{(2)}$.

We consider the same three-mode architecture as in the main text, consisting of two computational modes ($1$ and $2$) coupled via a tunable coupler ($c$), such that the indices $i, j \in \{1, 2, c\}$. In this system, the interaction leads to hybridization between the modes, such that each dressed operator acquires contributions from the others.

Using the expansion above, we express the dressed annihilation operators in terms of the bare modes. To second order in the coupling strengths, this yields the relaxation jump operators $\hat{L}_1^k(t) = \hat{a}_k(t)$:
\begin{subequations} \label{eq:dressed_relaxation}
\begin{align}
\hat{L}_1^{q_1}(t) &= \hat{a}_1 \mleft( 1 - \frac{g_{1c}^2}{2\Delta_{1c}^2} - \frac{g_{12}^2}{2\Delta_{12}^2} \mright) - \hat{a}_c \mleft( \frac{g_{1c}}{\Delta_{1c}} - \frac{g_{12} g_{2c}}{2 \Delta_{12} \Delta_{2c}} \mright) - \hat{a}_2 \mleft( \frac{g_{12}}{\Delta_{12}} + \frac{g_{1c} g_{2c}}{2 \Delta_{1c} \Delta_{2c}} \mright), \label{eq:L1_rel} \\
\hat{L}_1^{c}(t) &= \hat{a}_c \mleft( 1 - \frac{g_{1c}^2}{2\Delta_{1c}^2} - \frac{g_{2c}^2}{2\Delta_{2c}^2} \mright) + \hat{a}_1 \mleft( \frac{g_{1c}}{\Delta_{1c}} + \frac{g_{12} g_{2c}}{2 \Delta_{12} \Delta_{2c}} \mright) + \hat{a}_2 \mleft( \frac{g_{2c}}{\Delta_{2c}} - \frac{g_{1c} g_{12}}{2 \Delta_{1c} \Delta_{12}} \mright), \label{eq:Lc_rel} \\
\hat{L}_1^{q_2}(t) &= \hat{a}_2 \mleft( 1 - \frac{g_{12}^2}{2\Delta_{12}^2} - \frac{g_{2c}^2}{2\Delta_{2c}^2} \mright) - \hat{a}_c \mleft( \frac{g_{2c}}{\Delta_{2c}} + \frac{g_{1c} g_{12}}{2 \Delta_{1c} \Delta_{12}} \mright) + \hat{a}_1 \mleft( \frac{g_{12}}{\Delta_{12}} - \frac{g_{1c} g_{2c}}{2 \Delta_{1c} \Delta_{2c}} \mright). \label{eq:L2_rel}
\end{align}
\end{subequations}

For dephasing, the jump operators are given by $\hat{L}_\phi^{k} = \hat{a}_k^\dagger \hat{a}_k$. Using the same transformation as above, we express them in the dressed frame as
\begin{equation}
\hat{L}_\phi^{k}(t) = e^{\hat{S}} \hat{a}_k^\dagger \hat{a}_k e^{-\hat{S}} = \hat{a}_k^\dagger(t)\hat{a}_k(t),
\end{equation}
where we used the relation $e^{\hat{S}} \hat{a}_k^\dagger \hat{a}_k e^{-\hat{S}} = (e^{\hat{S}} \hat{a}_k^\dagger e^{-\hat{S}})(e^{\hat{S}} \hat{a}_k e^{-\hat{S}})$.

Substituting the expressions for the dressed annihilation operators and maintaining terms up to second order in the coupling strengths, we obtain the dressed dephasing operators. The resulting expressions contain diagonal contributions, proportional to $\hat{a}_i^\dagger \hat{a}_i$, reflecting how dephasing contributions are redistributed between the modes, and off-diagonal terms of the form $\hat{a}_i^\dagger \hat{a}_j$ ($i \neq j$), together with their Hermitian conjugates. This yields
\begin{subequations} \label{eq:dressed_dephasing}
\begin{align}
\hat{L}_\phi^{q_1}(t) &= \hat{L}_\phi^{q_1} \mleft( 1 - \frac{g_{1c}^2}{\Delta_{1c}^2} - \frac{g_{12}^2}{\Delta_{12}^2} \mright) + \hat{L}_\phi^{c} \frac{g_{1c}^2}{\Delta_{1c}^2} + \hat{L}_\phi^{q_2} \frac{g_{12}^2}{\Delta_{12}^2} \nonumber \\
&\quad - \mleft[ \hat{a}_1^\dagger \hat{a}_c \mleft( \frac{g_{1c}}{\Delta_{1c}} - \frac{g_{12} g_{2c}}{2 \Delta_{12} \Delta_{2c}} \mright) + \hat{a}_1^\dagger \hat{a}_2 \mleft( \frac{g_{12}}{\Delta_{12}} + \frac{g_{1c} g_{2c}}{2 \Delta_{1c} \Delta_{2c}} \mright) - \hat{a}_c^\dagger \hat{a}_2 \frac{g_{1c} g_{12}}{\Delta_{1c} \Delta_{12}} + \text{h.c.} \mright], \label{eq:L1_phi} \\
\hat{L}_\phi^{c}(t) &= \hat{L}_\phi^{c} \mleft( 1 - \frac{g_{1c}^2}{\Delta_{1c}^2} - \frac{g_{2c}^2}{\Delta_{2c}^2} \mright) + \hat{L}_\phi^{q_1} \frac{g_{1c}^2}{\Delta_{1c}^2} + \hat{L}_\phi^{q_2} \frac{g_{2c}^2}{\Delta_{2c}^2} \nonumber \\
&\quad + \mleft[ \hat{a}_c^\dagger \hat{a}_1 \mleft( \frac{g_{1c}}{\Delta_{1c}} + \frac{g_{12} g_{2c}}{2 \Delta_{12} \Delta_{2c}} \mright) + \hat{a}_c^\dagger \hat{a}_2 \mleft( \frac{g_{2c}}{\Delta_{2c}} - \frac{g_{1c} g_{12}}{2 \Delta_{1c} \Delta_{12}} \mright) + \hat{a}_1^\dagger \hat{a}_2 \frac{g_{1c} g_{2c}}{\Delta_{1c} \Delta_{2c}} + \text{h.c.} \mright], \label{eq:Lc_phi} \\
\hat{L}_\phi^{q_2}(t) &=\hat{L}_\phi^{q_2} \mleft( 1 - \frac{g_{12}^2}{\Delta_{12}^2} - \frac{g_{2c}^2}{\Delta_{2c}^2} \mright) + \hat{L}_\phi^{q_1} \frac{g_{12}^2}{\Delta_{12}^2} + \hat{L}_\phi^{c} \frac{g_{2c}^2}{\Delta_{2c}^2} \nonumber \\
&\quad + \mleft[ \hat{a}_2^\dagger \hat{a}_1 \mleft( \frac{g_{12}}{\Delta_{12}} - \frac{g_{1c} g_{2c}}{2 \Delta_{1c} \Delta_{2c}} \mright) - \hat{a}_2^\dagger \hat{a}_c \mleft( \frac{g_{2c}}{\Delta_{2c}} + \frac{g_{1c} g_{12}}{2 \Delta_{1c} \Delta_{12}} \mright) - \hat{a}_1^\dagger \hat{a}_c \frac{g_{12} g_{2c}}{\Delta_{12} \Delta_{2c}} + \text{h.c.} \mright]. \label{eq:L2_phi}
\end{align}
\end{subequations}

To evaluate the fidelity-reduction integrand in Eq.~(5) in the main text, we compute the trace over the computational subspace. For the present system, this subspace is given by
\begin{equation}
    \mathcal{B} = \{|00\rangle, |01\rangle, |10\rangle, |11\rangle\}\otimes |0\rangle_c,
\end{equation}
and we denote the corresponding partial trace by $\operatorname{Tr}_{\text{cmp}}(\cdot)$.
%
Because the trace is restricted to states with the coupler in its ground state, any expectation value involving an unpaired creation or annihilation operator of the coupler mode vanishes. Contributions involving excitations of the coupler are therefore excluded by the projection onto the computational subspace.
Using the complementary indicator $\bar{\delta}_{ic} \equiv 1 - \delta_{ic}$, we obtain the fundamental quadratic trace
\begin{equation} \label{eq:trace_quadratic}
\operatorname{Tr}_{\text{cmp}}\mleft(\hat{a}_i^\dagger \hat{a}_j\mright) = 2 \delta_{ij} \bar{\delta}_{ic}.
\end{equation}

For the bi-quadratic operator moments required to compute the fidelity reduction, one approach is to rearrange the product into normal order using the bosonic commutator $[\hat{a}_j, \hat{a}_k^\dagger] = \delta_{jk}$:
\begin{equation}
\operatorname{Tr}_{\text{cmp}}\mleft(\hat{a}_i^\dagger \hat{a}_j \hat{a}_k^\dagger \hat{a}_l\mright) = \operatorname{Tr}_{\text{cmp}}\mleft(\hat{a}_i^\dagger \hat{a}_k^\dagger \hat{a}_j \hat{a}_l\mright) + \delta_{jk} \operatorname{Tr}_{\text{cmp}}\mleft(\hat{a}_i^\dagger \hat{a}_l\mright).
\end{equation}

The second term follows directly from \eqref{eq:trace_quadratic}. For the normal-ordered contribution, the restriction to the computational subspace implies that only terms that remove and subsequently restore excitations contribute to the trace. In particular, terms with repeated annihilation on the same mode vanish.

The trace is therefore non-zero only when the annihilation operators act on distinct modes ($j \neq l$) and the creation operators restore those modes, such that $\{i,k\} = \{j,l\}$. Combining these conditions yields
\begin{align} \label{eq:trace_quartic_commutation}
\operatorname{Tr}_{\text{cmp}}\mleft(\hat{a}_i^\dagger \hat{a}_j \hat{a}_k^\dagger \hat{a}_l\mright) &= \mleft( \delta_{ij}\delta_{kl} + \delta_{il}\delta_{jk}\mright)\bar\delta_{jl} \bar{\delta}_{ic} \bar{\delta}_{jc} \bar{\delta}_{kc} \bar{\delta}_{lc}+ 2\delta_{jk}\delta_{il}\bar\delta_{ic}.
\end{align}

Inserting the computational identity operator from \eqref{I_cmp}, we write
\begin{equation} \label{eq:trace_quartic_expansion}
\operatorname{Tr}_{\text{cmp}}\mleft(\hat{a}_i^\dagger \hat{a}_j \hat{\mathds{1}}_{\text{cmp}} \hat{a}_k^\dagger \hat{a}_l\mright) = \sum_{n, m \in \mathcal{B}} \langle n | \hat{a}_i^\dagger \hat{a}_j | m \rangle \langle m | \hat{a}_k^\dagger \hat{a}_l | n \rangle.
\end{equation}
The restriction to the computational subspace implies that only transitions that remove and subsequently restore excitations contribute to the trace. Diagonal contributions arise for $i = j$ and $k = l$, while off-diagonal contributions require that the second operator reverses the first transition, which gives $j = k$ and $i = l$. In addition, all indices must correspond to computational modes, excluding the coupler. This yields
%
\begin{equation} \label{eq:trace_quartic_final}
\operatorname{Tr}_{\text{cmp}}\mleft(\hat{a}_i^\dagger \hat{a}_j \hat{\mathds{1}}_{\text{cmp}} \hat{a}_k^\dagger \hat{a}_l\mright) = \mleft( \delta_{ij}\delta_{kl}(2\delta_{jk}+\bar\delta_{jk}) +  \bar\delta_{ij}\bar\delta_{kl}\delta_{il}\delta_{jk} \mright) \bar{\delta}_{ic} \bar{\delta}_{jc} \bar{\delta}_{kc} \bar{\delta}_{lc}.
\end{equation}

Applying these trace identities, we obtain the effective relaxation and dephasing contributions to the fidelity reduction:
\begin{subequations} \label{eq:fidelity_rates}
\begin{align}
\delta F_\text{relaxation} &= -\frac{2}{5} \Gamma^{q_1}_1 \mleft( 1 - \frac{g_{1c}^2}{\Delta_{1c}^2} \mright) -\frac{2}{5} \Gamma^{c}_1 \mleft( \frac{g_{1c}^2}{\Delta_{1c}^2} + \frac{g_{2c}^2}{\Delta_{2c}^2} \mright) -\frac{2}{5} \Gamma^{q_2}_1 \mleft( 1 - \frac{g_{2c}^2}{\Delta_{2c}^2} \mright), \label{eq:deltaF_rel} \\
\delta F_\text{dephasing} &= -\frac{2}{5} \Gamma_\phi^{q_1} \mleft( 1 - \frac{g_{1c}^2}{2\Delta_{1c}^2} + \frac{5 g_{12}^2}{\Delta_{12}^2} \mright) -\Gamma_\phi^c \mleft( \frac{g_{1c}^2}{\Delta_{1c}^2} + \frac{g_{2c}^2}{\Delta_{2c}^2} \mright) -\frac{2}{5} \Gamma_\phi^{q_2} \mleft( 1 + \frac{5 g_{12}^2}{\Delta_{12}^2} + \frac{g_{2c}^2}{2\Delta_{2c}^2} \mright). \label{eq:deltaF_phi}
\end{align}
\end{subequations}
The total fidelity reduction is then given by
\begin{equation}
    \delta F = \delta F_\text{relaxation} + \delta F_\text{dephasing} ,
\end{equation}
which reproduces Eq.~(16) in the main text.


\section{Leakage} \label{sc:leakge}

We choose two average leakage metrics to quantify the population lost from the computational subspace during the gate operation. The total leakage, representing the average population lost from the computational subspace, is given by~\cite{wood2018}
\begin{equation}
    \Lambda_{\text{total}}(\mathcal{E}) = 1 - \text{Tr}\mleft[ \hat{\mathds{1}}_{\mathrm{cmp}} \, \mathcal{E}\mleft(\frac{\hat{\mathds{1}}_{\mathrm{cmp}}}{d}\mright) \mright],
\end{equation}
where $\hat{\mathds{1}}_{\mathrm{cmp}}$ is the projector onto the $d$-dimensional computational subspace [see \eqref{I_cmp}]. Additionally, we define the state-specific leakage into a particular non-computational state $\ket{m}$, with the corresponding projector $\Pi_m = \ket{m}\bra{m}$:
\begin{equation}
    \Lambda_{m}(\mathcal{E}) = \text{Tr}\mleft[ \Pi_m \, \mathcal{E}\mleft(\frac{\hat{\mathds{1}}_{\mathrm{cmp}}}{d}\mright) \mright].
\end{equation}

To evaluate these metrics, we utilize the first-order Dyson expansion of the time-evolution superoperator, $\mathcal{U}_L(\tau,0) \approx \mathcal{U}^{(0)}_L(\tau,0) + \mathcal{U}^{(1)}_L(\tau,0)$, given by \eqref{eq:1st o expansion}. The zeroth-order term $\mathcal{U}^{(0)}_L$ represents the ideal adiabatic evolution, where the unitary operator can be decomposed as $\hat{U}(t) = \hat{U}_\varphi(t)\hat{U}_\psi(t)$. Here, $\hat{U}_\varphi(t)$ contains the dynamical and geometric phases, and $\hat{U}_\psi(t)$ is the frame-transformation operator that maps the initial eigenstates to the instantaneous eigenstates.

To evaluate traces at the end of the gate, we substitute the explicit unitary action into the trace of an observable $\hat{O}(\tau)$:
\begin{equation}
    \Tr\mleft[ \hat{O}(\tau) \, \mathcal{U}^{(1)}_L(\tau,0)\rho \mright] = \sum_{k=1}^{N_L} \int_{0}^{\tau} dt' \Gamma_k(t') \Tr\mleft[ \hat{O}(\tau) \, \hat{U}(\tau,t') \mathcal{D}[\hat{L}_k]\mleft(\hat{U}(t',0) \rho \hat{U}^\dagger(t',0)\mright) \hat{U}^\dagger(\tau,t') \mright].
\end{equation}

We explicitly use the cyclic property of the trace to unwrap the unitary evolution from the state and move it onto the observable:
\begin{align}
    \Tr&\mleft[ \hat{O}(\tau) \, \hat{U}(\tau,t') \mathcal{D}[\hat{L}_k](\rho(t')) \hat{U}^\dagger(\tau,t') \mright] = \Tr\mleft[ \mleft( \hat{U}^\dagger(\tau,t') \hat{O}(\tau) \hat{U}(\tau,t') \mright) \mathcal{D}[\hat{L}_k](\rho(t')) \mright].
\end{align}

While the cyclic loop $\ket{m(\tau)} = \ket{m(0)}$ holds for the adiabatic eigenstates themselves, it does not hold for the accumulated phase. However, because our observables $\hat{O}$ are formed by eigenstate projectors (e.g., $\sum_m\Pi_m(\tau) = \sum_m\ket{m(\tau)}\bra{m(\tau)}$), the backward evolution from $\tau$ to $t'$ yields complex phases $e^{-i(\varphi_m(\tau)-\varphi_m(t'))}$ that strictly cancel in the outer product:
\begin{equation}
    \hat{U}^\dagger(\tau,t') \ket{m(\tau)}\bra{m(\tau)} \hat{U}(\tau,t') = \ket{m(t')}\bra{m(t')} = \hat{U}_\psi(t') \ket{m(0)}\bra{m(0)} \hat{U}_\psi^\dagger(t').
\end{equation}
This explicitly shows that the backward-evolved observable is simply the frame-transformed initial observable $\hat{O}(t') = \hat{U}_\psi(t') \hat{O}(0) \hat{U}_\psi^\dagger(t')$, fully independent of the phase operator $\hat{U}_\varphi(t)$.

Similarly, because the initial state $\rho = \frac{\mathds{1}_{\rm cmp}}{d}$~\cite{wood2018} is diagonal in the initial eigenbasis, its forward evolution $\rho(t') = \hat{U}(t',0) \rho \hat{U}^\dagger(t',0)$ perfectly cancels the phase operators:
\begin{equation}
    \rho(t') = \hat{U}_\psi(t') \rho \hat{U}_\psi^\dagger(t').
\end{equation}

Substituting these directly into the trace and applying the cyclic property once more yields
\begin{align}
    \Tr&\mleft[ \mleft( \hat{U}_\psi(t') \hat{O}(0) \hat{U}_\psi^\dagger(t') \mright) \mathcal{D}[\hat{L}_k]\mleft(\hat{U}_\psi(t') \rho \hat{U}_\psi^\dagger(t')\mright) \mright] = \Tr\mleft[ \hat{O}(0) \, \mathcal{D}\mleft[\hat{U}_\psi^\dagger(t')\hat{L}_k\hat{U}_\psi(t')\mright](\rho) \mright].
\end{align}
So we can absorb all time dependence strictly via the frame-transformation operator to define the dressed jump operator, just as in the main text:
\begin{equation}
    \hat{L}_k(t') = \hat{U}_\psi^\dagger(t') \hat{L}_k \hat{U}_\psi(t').
\end{equation}

We first compute the first-order leakage contribution into a specific target state $\ket{m}$ outside the computational subspace. Because the ideal unitary perfectly preserves the computational subspace, the zeroth-order term evaluates to zero, and the leakage is entirely driven by $\mathcal{U}^{(1)}_L$. Using the dressed jump operators, the state-specific leakage is
\begin{equation}
    \Lambda_{m}^{(1)} = \frac{1}{d} \sum_{k=1}^{N_L} \int_0^\tau dt' \Gamma_k(t') \text{Tr} \mleft[ \Pi_m \mleft( \hat{L}_k(t') \hat{\mathds{1}}_{\mathrm{cmp}} \hat{L}_k^\dagger(t') - \frac{1}{2}\mleft\{\hat{L}_k^\dagger(t')\hat{L}_k(t'), \hat{\mathds{1}}_{\mathrm{cmp}}\mright\} \mright) \mright].
\end{equation}
Since $\ket{m}$ is strictly orthogonal to the computational subspace ($\Pi_m \hat{\mathds{1}}_{\mathrm{cmp}} = 0$), the anticommutator terms vanish. Expanding $\hat{\mathds{1}}_{\mathrm{cmp}} = \sum_{i \in \mathrm{cmp}} \ket{i}\bra{i}$, we obtain the closed-form expression
\begin{equation}
    \Lambda_{m}^{(1)} = \frac{1}{d} \sum_{k=1}^{N_L} \int_0^\tau dt' \Gamma_k(t') \sum_{i \in \mathrm{cmp}} \mleft| \bra{m} \hat{L}_k(t') \ket{i} \mright|^2.
\end{equation}

Similarly, the first-order total leakage out of the computational subspace is given by $\Lambda_{\text{total}}^{(1)} = - \Tr[ \mathds{1}_{\rm cmp} \, \mathcal{U}^{(1)}_L(\tau,0)\frac{\mathds{1}_{\rm cmp}}{d}]$. Expanding the Lindblad dissipator and utilizing the cyclic property of the trace, the anticommutator traces can be combined, yielding
\begin{equation}
    \Lambda_{\text{total}}^{(1)} = \frac{1}{d} \sum_{k=1}^{N_L} \int_0^\tau dt' \Gamma_k(t') \mleft( \Tr_{\text{cmp}}\mleft[\hat{L}_k^\dagger(t') \hat{L}_k(t') \mright] - \Tr_{\text{cmp}}\mleft[\hat{L}_k^\dagger(t') \mathds{1}_{\rm cmp} \hat{L}_k(t') \mright] \mright).
\end{equation}

From these expressions, we extract the explicit time-dependent leakage rates $\dot{\Lambda}_{\to \ket{m}}(t)$ for the specific transitions in our system:
\begin{subequations}
\begin{align}
\dot{\Lambda}_{\to \ket{001}} &= \Gamma_{\phi}^{q1} \mleft( \frac{g_{1c}^2}{2 \Delta_{1c}^2} \mright) + \Gamma_{\phi}^{q2} \mleft( \frac{g_{2c}^2}{2 \Delta_{2c}^2} \mright) + \Gamma_{\phi}^{c} \mleft( \frac{g_{1c}^2}{2 \Delta_{1c}^2}+\frac{g_{2c}^2}{2 \Delta_{2c}^2} \mright) , \\
\dot{\Lambda}_{\to \ket{002}} &= 0, \\
\dot{\Lambda}_{\to \ket{011}} &= \Gamma_{\phi}^{q1} \mleft( \frac{g_{1c}^2}{2 \Delta_{1c}^2} \mright) + \Gamma_{\phi}^{c} \mleft( \frac{g_{1c}^2}{2 \Delta_{1c}^2} \mright) , \\
\dot{\Lambda}_{\to \ket{101}} &= \Gamma_{\phi}^{q2} \mleft( \frac{g_{2c}^2}{2 \Delta_{2c}^2} \mright) + \Gamma_{\phi}^{c} \mleft( \frac{g_{2c}^2}{2 \Delta_{2c}^2} \mright) , \\
\dot{\Lambda}_{\to \ket{020}} &= \Gamma_{\phi}^{q1} \mleft( \frac{g_{12}^2}{\mleft(\alpha_2-\Delta_{12}\mright)^2} \mright) + \Gamma_{\phi}^{q2} \mleft( \frac{g_{12}^2}{\mleft(\alpha_2-\Delta_{12}\mright)^2} \mright) , \\
\dot{\Lambda}_{\to \ket{200}} &= \Gamma_{\phi}^{q1} \mleft( \frac{g_{12}^2}{\mleft(\alpha_1+\Delta_{12}\mright)^2} \mright) + \Gamma_{\phi}^{q2} \mleft( \frac{g_{12}^2}{\mleft(\alpha_1+\Delta_{12}\mright)^2} \mright) .
\end{align}
\end{subequations}
The total leakage rate $\dot{\Lambda}_{\text{total}}(t)$ is the sum of these state-specific rates:
\begin{align}
\dot{\Lambda}_{\text{total}} &=
\Gamma_{\phi}^{q1} \mleft( \frac{g_{1c}^2}{\Delta_{1c}^2}+\frac{g_{12}^2}{\mleft(\alpha_2-\Delta_{12}\mright)^2}+\frac{g_{12}^2}{\mleft(\alpha_1+\Delta_{12}\mright)^2} \mright) \nonumber \\
&\quad + \Gamma_{\phi}^{q2} \mleft( \frac{g_{2c}^2}{\Delta_{2c}^2}+\frac{g_{12}^2}{\mleft(\alpha_2-\Delta_{12}\mright)^2}+\frac{g_{12}^2}{\mleft(\alpha_1+\Delta_{12}\mright)^2} \mright) \nonumber \\
&\quad + \Gamma_{\phi}^{c} \mleft( \frac{g_{1c}^2}{\Delta_{1c}^2}+\frac{g_{2c}^2}{\Delta_{2c}^2} \mright) .
\end{align}

As seen from these final explicit expressions, the accumulated leakage depends exclusively on the pure dephasing rates.

\section{Fourth-order analytical expression for fidelity reduction with anharmonicity} \label{anyl_exp}

To obtain the fourth-order expression with anharmonicities for the fidelity-reduction rate of the adiabatic CZ gate in our example, generalizing Eq.~(16) in the main text, we construct an explicit matrix representation of the system's Hamiltonian [Eq.~(15) in the main text] using the rotating-wave approximation. We impose a level truncation, restricting each subsystem to the lowest $N=3$ energy levels.
The local annihilation operators are then defined as $3 \times 3$ matrices in the truncated Fock basis:
\begin{align}
a = \begin{pmatrix} 
0 & 1 & 0 \\ 
0 & 0 & \sqrt{2} \\ 
0 & 0 & 0 
\end{pmatrix} .   
\end{align}

We construct the total unperturbed Hamiltonian $H_0$ and the perturbation $V$ in the full $3^3 = 27$-dimensional joint Hilbert space using the tensor product ($\otimes$). For example, the operator for the first subsystem is expanded as $a_1 = a \otimes \mathbb{I} \otimes \mathbb{I}$, where $\mathbb{I}$ is the $3 \times 3$ identity matrix. This exact matrix representation allows for the direct numerical evaluation of commutators.

We employ an iterative approach to eliminate off-diagonal interactions order by order. The first-order generator $S_1$ is constructed to eliminate the first-order perturbation $V$. In the eigenbasis of $H_0$, the matrix elements of $S_1$ are found by solving $[H_0, S_1] = -V$, which yields
\begin{equation}
(S_1)_{ij} = \frac{V_{ij}}{E_i - E_j} ,
\end{equation}
where $E_i$ are the eigenvalues of $H_0$.

Applying this transformation generates new second-order terms. The second-order effective Hamiltonian components are computed via $H_2 = \frac{1}{2}[S_1, V]$. We separate $H_2$ into its diagonal part $D_2$ and off-diagonal part $V_2 = H_2 - D_2$. To eliminate these residual second-order off-diagonal transitions, we compute a second generator $S_2$ such that $[H_0, S_2] = -V_2$, following the same element-wise evaluation:
\begin{equation}
(S_2)_{ij} = \frac{(V_2)_{ij}}{E_i - E_j} .
\end{equation}

To find the jump operators in the dressed frame, the fundamental operators (e.g., $a_1, a_2, a_c$) must be transformed using the same unitary matrices. We utilize the BCH lemma and truncate the BCH expansion at the fourth order (i.e., up to four nested commutators). The transformation is applied iteratively: we first compute the first-order dressed operators $\tilde{O}^{(1)} = e^{S_1} O e^{-S_1}$ truncated at order four, and subsequently apply the second-order generator to obtain the fully dressed operators $\tilde{O}^{(2)} = e^{S_2} \tilde{O}^{(1)} e^{-S_2}$, again utilizing a fourth-order BCH expansion.

Finally, by substituting these fourth-order dressed operators into the trace evaluation in Eq.~(5) in the main text, we obtain the fidelity-reduction integrand as:

\begin{align}
&\delta F^{\text{relaxation}} = \nonumber \\
& -\frac{2}{5}\Gamma_{1}^{q1} \mleft( \begin{aligned}
& 1 -\frac{g_{1c}^2}{\Delta _{1c}^2}-\frac{g_{12}^2}{\mleft(\alpha _2-\Delta _{12}\mright)^2}+\frac{g_{12}^2}{\mleft(\alpha _1+\Delta _{12}\mright)^2}  + \frac{g_{12} g_{1c} g_{2c}}{\Delta _{2c} \mleft(\alpha _1+\Delta _{12}\mright) \mleft(\Delta _{1c}+\alpha _1\mright)} \\
& + \frac{2 g_{12}^2 g_{1c}^2}{3 \Delta _{1c}^2 \mleft(\alpha _2-\Delta _{12}\mright)^2}+\frac{11 g_{12}^2 g_{1c}^2}{24 \Delta _{12}^2 \Delta _{1c}^2}+\frac{g_{1c}^4}{3 \Delta _{1c}^4}+\frac{2 g_{12}^4}{3 \mleft(\alpha _2-\Delta _{12}\mright)^4}-\frac{2 g_{12}^4}{3 \mleft(\alpha _1+\Delta _{12}\mright)^4} \\
& + \frac{g_{12}^2 g_{1c}^2}{4 \Delta _{12} \Delta _{1c} \mleft(\alpha _1+\Delta _{12}\mright) \mleft(\Delta _{1c}+\alpha _1\mright)}-\frac{g_{12}^2 g_{1c}^2}{2 \Delta _{1c} \mleft(\alpha _1+\Delta _{12}\mright)^2 \mleft(\Delta _{1c}+\alpha _1\mright)}-\frac{g_{12}^2 g_{1c}^2}{2 \Delta _{12} \Delta _{1c}^2 \mleft(\alpha _1+\Delta _{12}\mright)}\\ 
&+\frac{g_{12}^2 g_{1c}^2}{2 \Delta _{1c}^2 \mleft(\alpha _1+\Delta _{12}\mright)^2}-\frac{13 g_{12}^2 g_{1c}^2}{24 \mleft(\alpha _1+\Delta _{12}\mright)^2 \mleft(\Delta _{1c}+\alpha _1\mright)^2} + \frac{g_{1c}^2 g_{2c}^2}{12 \Delta _{1c}^2 \mleft(\alpha _{\text{c}}-\Delta _{2c}\mright)^2}\\  
& +\frac{g_{1c}^2 g_{2c}^2}{12 \Delta _{1c}^2 \mleft(-\Delta _{2c}-\alpha _2\mright)^2}+\frac{g_{12}^2 g_{2c}^2}{3 \Delta _{2c}^2 \mleft(\alpha _2-\Delta _{12}\mright)^2}-\frac{g_{12}^2 g_{2c}^2}{3 \Delta _{2c}^2 \mleft(\alpha _1+\Delta _{12}\mright)^2}-\frac{g_{12}^2 g_{2c}^2}{4 \Delta _{12}^2 \Delta _{2c}^2} \\
& + \frac{g_{1c}^2 g_{2c}^2}{6 \Delta _{1c} \Delta _{2c} \mleft(\Delta _{1c}-\alpha _{\text{c}}\mright) \mleft(\alpha _{\text{c}}-\Delta _{2c}\mright)}-\frac{g_{1c}^2 g_{2c}^2}{4 \Delta _{2c}^2 \mleft(\alpha _{\text{c}}-\Delta _{1c}\mright)^2}+\frac{g_{1c}^2 g_{2c}^2}{4 \Delta _{2c}^2 \mleft(\Delta _{1c}+\alpha _1\mright)^2}+\frac{g_{1c}^2 g_{2c}^2}{3 \Delta _{1c}^2 \Delta _{2c}^2}\\
&+\frac{g_{12}^2 g_{2c}^2}{6 \mleft(\alpha _2-\Delta _{12}\mright)^2 \mleft(\Delta _{2c}+\alpha _2\mright)^2} -\frac{g_{12}^2 g_{2c}^2}{6 \Delta _{12} \Delta _{2c} \mleft(\Delta _{12}-\alpha _2\mright) \mleft(\Delta _{2c}+\alpha _2\mright)}
\end{aligned} \mright) \nonumber \\
&-\frac{2}{5} \Gamma_{1}^{q2}\mleft( \begin{aligned}
& 1 -\frac{g_{2c}^2}{\Delta _{2c}^2}+\frac{g_{12}^2}{\mleft(\alpha _2-\Delta _{12}\mright)^2}-\frac{g_{12}^2}{\mleft(\alpha _1+\Delta _{12}\mright)^2}  + \frac{g_{12} g_{1c} g_{2c}}{\Delta _{1c} \mleft(\alpha _2-\Delta _{12}\mright) \mleft(\Delta _{2c}+\alpha _2\mright)} \\
& -\frac{g_{12}^2 g_{1c}^2}{3 \Delta _{1c}^2 \mleft(\alpha _2-\Delta _{12}\mright)^2}+\frac{g_{12}^2 g_{1c}^2}{3 \Delta _{1c}^2 \mleft(\alpha _1+\Delta _{12}\mright)^2}-\frac{g_{12}^2 g_{1c}^2}{4 \Delta _{12}^2 \Delta _{1c}^2}-\frac{2 g_{12}^4}{3 \mleft(\alpha _2-\Delta _{12}\mright)^4}+\frac{2 g_{12}^4}{3 \mleft(\alpha _1+\Delta _{12}\mright)^4} \\
& -\frac{g_{1c}^2 g_{2c}^2}{4 \Delta _{1c}^2 \mleft(\alpha _{\text{c}}-\Delta _{2c}\mright)^2}+\frac{g_{1c}^2 g_{2c}^2}{4 \Delta _{1c}^2 \mleft(-\Delta _{2c}-\alpha _2\mright)^2}-\frac{g_{12}^2 g_{1c}^2}{6 \Delta _{12} \Delta _{1c} \mleft(\alpha _1+\Delta _{12}\mright) \mleft(\Delta _{1c}+\alpha _1\mright)} \\
& +\frac{g_{12}^2 g_{1c}^2}{6 \mleft(\alpha _1+\Delta _{12}\mright)^2 \mleft(\Delta _{1c}+\alpha _1\mright)^2}+\frac{g_{2c}^4}{3 \Delta _{2c}^4} + \frac{g_{1c}^2 g_{2c}^2}{12 \Delta _{2c}^2 \mleft(\alpha _{\text{c}}-\Delta _{1c}\mright)^2}-\frac{g_{12}^2 g_{2c}^2}{2 \Delta _{12} \Delta _{2c}^2 \mleft(\Delta _{12}-\alpha _2\mright)}\\ 
&+\frac{g_{12}^2 g_{2c}^2}{2 \Delta _{2c}^2 \mleft(\alpha _2-\Delta _{12}\mright)^2}+\frac{2 g_{12}^2 g_{2c}^2}{3 \Delta _{2c}^2 \mleft(\alpha _1+\Delta _{12}\mright)^2}+\frac{11 g_{12}^2 g_{2c}^2}{24 \Delta _{12}^2 \Delta _{2c}^2} + \frac{g_{1c}^2 g_{2c}^2}{6 \Delta _{1c} \Delta _{2c} \mleft(\Delta _{1c}-\alpha _{\text{c}}\mright) \mleft(\alpha _{\text{c}}-\Delta _{2c}\mright)} \\ 
&+\frac{g_{1c}^2 g_{2c}^2}{12 \Delta _{2c}^2 \mleft(\Delta _{1c}+\alpha _1\mright)^2}+\frac{g_{1c}^2 g_{2c}^2}{3 \Delta _{1c}^2 \Delta _{2c}^2}-\frac{g_{12}^2 g_{2c}^2}{2 \Delta _{2c} \mleft(\alpha _2-\Delta _{12}\mright)^2 \mleft(\Delta _{2c}+\alpha _2\mright)} \\ & -\frac{13 g_{12}^2 g_{2c}^2}{24 \mleft(\alpha _2-\Delta _{12}\mright)^2 \mleft(\Delta _{2c}+\alpha _2\mright)^2} + \frac{g_{12}^2 g_{2c}^2}{4 \Delta _{12} \Delta _{2c} \mleft(\Delta _{12}-\alpha _2\mright) \mleft(\Delta _{2c}+\alpha _2\mright)}
\end{aligned} \mright)\nonumber \\
& -\frac{2}{5}\Gamma_{1}^{c} \mleft( \begin{aligned}
& \frac{g_{1c}^2}{\Delta _{1c}^2}+\frac{g_{2c}^2}{\Delta _{2c}^2} -\frac{g_{12} g_{1c} g_{2c}}{\Delta _{2c} \mleft(\alpha _1+\Delta _{12}\mright) \mleft(\Delta _{1c}+\alpha _1\mright)}-\frac{g_{12} g_{1c} g_{2c}}{\Delta _{1c} \mleft(\alpha _2-\Delta _{12}\mright) \mleft(\Delta _{2c}+\alpha _2\mright)} \\
& -\frac{g_{12}^2 g_{1c}^2}{3 \Delta _{1c}^2 \mleft(\alpha _2-\Delta _{12}\mright)^2}-\frac{g_{12}^2 g_{1c}^2}{3 \Delta _{1c}^2 \mleft(\alpha _1+\Delta _{12}\mright)^2}+\frac{g_{12}^2 g_{1c}^2}{2 \mleft(\alpha _1+\Delta _{12}\mright)^2 \mleft(\Delta _{1c}+\alpha _1\mright)^2}-\frac{g_{1c}^4}{3 \Delta _{1c}^4} \\
&-\frac{g_{12}^2 g_{1c}^2}{12 \Delta _{12}^2 \Delta _{1c}^2} + \frac{7 g_{1c}^2 g_{2c}^2}{24 \Delta _{1c}^2 \mleft(\alpha _{\text{c}}-\Delta _{2c}\mright)^2}-\frac{g_{1c}^2 g_{2c}^2}{3 \Delta _{1c}^2 \mleft(-\Delta _{2c}-\alpha _2\mright)^2}+\frac{g_{12}^2 g_{1c}^2}{6 \Delta _{12} \Delta _{1c} \mleft(\alpha _1+\Delta _{12}\mright) \mleft(\Delta _{1c}+\alpha _1\mright)} \\ 
&-\frac{g_{12}^2 g_{2c}^2}{3 \Delta _{2c}^2 \mleft(\alpha _2-\Delta _{12}\mright)^2}-\frac{g_{2c}^4}{3 \Delta _{2c}^4}  + \frac{7 g_{1c}^2 g_{2c}^2}{24 \Delta _{2c}^2 \mleft(\alpha _{\text{c}}-\Delta _{1c}\mright)^2}-\frac{g_{1c}^2 g_{2c}^2}{3 \Delta _{2c}^2 \mleft(\Delta _{1c}+\alpha _1\mright)^2}-\frac{g_{1c}^2 g_{2c}^2}{6 \Delta _{1c}^2 \Delta _{2c}^2}\\
&-\frac{g_{12}^2 g_{2c}^2}{3 \Delta _{2c}^2 \mleft(\alpha _1+\Delta _{12}\mright)^2}-\frac{g_{12}^2 g_{2c}^2}{12 \Delta _{12}^2 \Delta _{2c}^2} -\frac{7 g_{1c}^2 g_{2c}^2}{12 \Delta _{1c} \Delta _{2c} \mleft(\Delta _{1c}-\alpha_{\text{c}}\mright) \mleft(\alpha_{\text{c}}-\Delta _{2c}\mright)}+\frac{g_{1c}^2 g_{2c}^2}{2 \Delta _{1c}^2 \Delta _{2c} \mleft(\alpha _{\text{c}}-\Delta _{2c}\mright)}\\
& -\frac{g_{1c}^2 g_{2c}^2}{2 \Delta _{1c} \Delta _{2c}^2 \mleft(\Delta _{1c}-\alpha _{\text{c}}\mright)} +\frac{g_{12}^2 g_{2c}^2}{6 \Delta _{12} \Delta _{2c} \mleft(\Delta _{12}-\alpha _2\mright) \mleft(\Delta _{2c}+\alpha _2\mright)} +\frac{g_{12}^2 g_{2c}^2}{2 \mleft(\alpha _2-\Delta _{12}\mright)^2 \mleft(\Delta _{2c}+\alpha _2\mright)^2}
\end{aligned} \mright) ,
\end{align}

\begin{align}
&\delta F^{\text{dephasing}} = \nonumber \\
& -\frac{2}{5}\Gamma_{\phi}^{q1} \mleft( \begin{aligned}
& 1 + \frac{g_{1c}^2}{2 \Delta _{1c}^2}+\frac{g_{12}^2}{2 \mleft(\alpha _2-\Delta _{12}\mright)^2}+\frac{9 g_{12}^2}{2 \mleft(\alpha _1+\Delta _{12}\mright)^2} + \frac{9 g_{12} g_{1c} g_{2c}}{2 \Delta _{2c} \mleft(\alpha _1+\Delta _{12}\mright) \mleft(\Delta _{1c}+\alpha _1\mright)} \\
& -\frac{10 g_{12}^2 g_{1c}^2}{3 \Delta _{1c}^2 \mleft(\alpha _2-\Delta _{12}\mright)^2}-\frac{5 g_{1c}^4}{3 \Delta _{1c}^4}+\frac{2 g_{12}^4}{3 \mleft(\alpha _1+\Delta _{12}\mright)^2 \mleft(\Delta _{12}-\alpha _2\mright)^2}-\frac{7 g_{12}^4}{3 \mleft(\alpha _2-\Delta _{12}\mright)^4}-\frac{5 g_{12}^4}{\mleft(\alpha _1+\Delta _{12}\mright)^4} \\
&  -\frac{g_{1c}^2 g_{2c}^2}{24 \Delta _{1c}^2 \mleft(-\Delta _{2c}-\alpha _2\mright)^2}-\frac{5 g_{12}^2 g_{1c}^2}{3 \Delta _{12} \Delta _{1c} \mleft(\alpha _1+\Delta _{12}\mright) \mleft(\Delta _{1c}+\alpha _1\mright)}+\frac{4 g_{12}^2 g_{1c}^2}{3 \Delta _{1c}^2 \mleft(\alpha _1+\Delta _{12}\mright)^2} \\ & -\frac{3 g_{12}^2 g_{1c}^2}{\mleft(\alpha _1+\Delta _{12}\mright)^2 \mleft(\Delta _{1c}+\alpha _1\mright)^2}-\frac{g_{12}^2 g_{1c}^2}{6 \Delta _{12}^2 \Delta _{1c}^2}  -\frac{g_{1c}^2 g_{2c}^2}{24 \Delta _{1c}^2 \mleft(\alpha _{\text{c}}-\Delta _{2c}\mright)^2}+\frac{g_{1c}^2 g_{2c}^2}{8 \Delta _{2c}^2 \mleft(\alpha _{\text{c}}-\Delta _{1c}\mright)^2} \\
&  -\frac{g_{12}^2 g_{2c}^2}{6 \Delta _{2c}^2 \mleft(\alpha _2-\Delta _{12}\mright)^2}-\frac{3 g_{12}^2 g_{2c}^2}{2 \Delta _{2c}^2 \mleft(\alpha _1+\Delta _{12}\mright)^2}+\frac{g_{12}^2 g_{2c}^2}{8 \Delta _{12}^2 \Delta _{2c}^2} -\frac{g_{1c}^2 g_{2c}^2}{12 \Delta _{1c} \Delta _{2c} \mleft(\Delta _{1c}-\alpha _{\text{c}}\mright) \mleft(\alpha _{\text{c}}-\Delta _{2c}\mright)}  \\
&  +\frac{9 g_{1c}^2 g_{2c}^2}{8 \Delta _{2c}^2 \mleft(\Delta _{1c}+\alpha _1\mright)^2}-\frac{g_{1c}^2 g_{2c}^2}{6 \Delta _{1c}^2 \Delta _{2c}^2}+\frac{g_{12}^2 g_{2c}^2}{12 \Delta _{12} \Delta _{2c} \mleft(\Delta _{12}-\alpha _2\mright) \mleft(\Delta _{2c}+\alpha _2\mright)}-\frac{g_{12}^2 g_{2c}^2}{12 \mleft(\alpha _2-\Delta _{12}\mright)^2 \mleft(\Delta _{2c}+\alpha _2\mright)^2}
\end{aligned} \mright) \nonumber \\
&-\frac{2}{5} \Gamma_{\phi}^{q2}\mleft( \begin{aligned}
& 1  + \frac{g_{2c}^2}{2 \Delta _{2c}^2}+\frac{9 g_{12}^2}{2 \mleft(\alpha _2-\Delta _{12}\mright)^2}+\frac{g_{12}^2}{2 \mleft(\alpha _1+\Delta _{12}\mright)^2} + \frac{9 g_{12} g_{1c} g_{2c}}{2 \Delta _{1c} \mleft(\alpha _2-\Delta _{12}\mright) \mleft(\Delta _{2c}+\alpha _2\mright)} \\
& -\frac{3 g_{12}^2 g_{1c}^2}{2 \Delta _{1c}^2 \mleft(\alpha _2-\Delta _{12}\mright)^2}+\frac{g_{12}^2 g_{1c}^2}{8 \Delta _{12}^2 \Delta _{1c}^2}+\frac{2 g_{12}^4}{3 \mleft(\alpha _1+\Delta _{12}\mright)^2 \mleft(\Delta _{12}-\alpha _2\mright)^2}-\frac{5 g_{12}^4}{\mleft(\alpha _2-\Delta _{12}\mright)^4}-\frac{7 g_{12}^4}{3 \mleft(\alpha _1+\Delta _{12}\mright)^4} \\
& + \frac{g_{1c}^2 g_{2c}^2}{8 \Delta _{1c}^2 \mleft(\alpha _{\text{c}}-\Delta _{2c}\mright)^2}+\frac{9 g_{1c}^2 g_{2c}^2}{8 \Delta _{1c}^2 \mleft(-\Delta _{2c}-\alpha _2\mright)^2}+\frac{g_{12}^2 g_{1c}^2}{12 \Delta _{12} \Delta _{1c} \mleft(\alpha _1+\Delta _{12}\mright) \mleft(\Delta _{1c}+\alpha _1\mright)}-\frac{g_{12}^2 g_{1c}^2}{6 \Delta _{1c}^2 \mleft(\alpha _1+\Delta _{12}\mright)^2} \\
&   -\frac{g_{12}^2 g_{1c}^2}{12 \mleft(\alpha _1+\Delta _{12}\mright)^2 \mleft(\Delta _{1c}+\alpha _1\mright)^2}-\frac{g_{1c}^2 g_{2c}^2}{24 \Delta _{2c}^2 \mleft(\alpha _{\text{c}}-\Delta _{1c}\mright)^2}+\frac{4 g_{12}^2 g_{2c}^2}{3 \Delta _{2c}^2 \mleft(\alpha _2-\Delta _{12}\mright)^2}-\frac{10 g_{12}^2 g_{2c}^2}{3 \Delta _{2c}^2 \mleft(\alpha _1+\Delta _{12}\mright)^2} \\
& -\frac{5 g_{2c}^4}{3 \Delta _{2c}^4} -\frac{g_{12}^2 g_{2c}^2}{6 \Delta _{12}^2 \Delta _{2c}^2}  -\frac{g_{1c}^2 g_{2c}^2}{12 \Delta _{1c} \Delta _{2c} \mleft(\Delta _{1c}-\alpha _{\text{c}}\mright) \mleft(\alpha _{\text{c}}-\Delta _{2c}\mright)}-\frac{g_{1c}^2 g_{2c}^2}{24 \Delta _{2c}^2 \mleft(\Delta _{1c}+\alpha _1\mright)^2}-\frac{g_{1c}^2 g_{2c}^2}{6 \Delta _{1c}^2 \Delta _{2c}^2}\\ 
&-\frac{5 g_{12}^2 g_{2c}^2}{3 \Delta _{12} \Delta _{2c} \mleft(\Delta _{12}-\alpha _2\mright) \mleft(\Delta _{2c}+\alpha _2\mright)}-\frac{3 g_{12}^2 g_{2c}^2}{\mleft(\alpha _2-\Delta _{12}\mright)^2 \mleft(\Delta _{2c}+\alpha _2\mright)^2}
\end{aligned} \mright)\nonumber \\
& -\frac{2}{5}\Gamma_{\phi}^{c} \mleft( \begin{aligned}
& \frac{5 g_{1c}^2}{2 \Delta _{1c}^2}+\frac{5 g_{2c}^2}{2 \Delta _{2c}^2}  -\frac{5 g_{12} g_{1c} g_{2c}}{2 \Delta _{2c} \mleft(\alpha _1+\Delta _{12}\mright) \mleft(\Delta _{1c}+\alpha _1\mright)}-\frac{5 g_{12} g_{1c} g_{2c}}{2 \Delta _{1c} \mleft(\alpha _2-\Delta _{12}\mright) \mleft(\Delta _{2c}+\alpha _2\mright)} \\
& -\frac{5 g_{12}^2 g_{1c}^2}{6 \Delta _{1c}^2 \mleft(\alpha _2-\Delta _{12}\mright)^2}-\frac{5 g_{12}^2 g_{1c}^2}{6 \Delta _{1c}^2 \mleft(\alpha _1+\Delta _{12}\mright)^2}+\frac{5 g_{12}^2 g_{1c}^2}{4 \mleft(\alpha _1+\Delta _{12}\mright)^2 \mleft(\Delta _{1c}+\alpha _1\mright)^2}-\frac{7 g_{1c}^4}{3 \Delta _{1c}^4}-\frac{5 g_{12}^2 g_{1c}^2}{24 \Delta _{12}^2 \Delta _{1c}^2} \\
&  + \frac{5 g_{1c}^2 g_{2c}^2}{3 \Delta _{1c}^2 \mleft(\alpha _{\text{c}}-\Delta _{2c}\mright)^2}-\frac{5 g_{1c}^2 g_{2c}^2}{6 \Delta _{1c}^2 \mleft(-\Delta _{2c}-\alpha _2\mright)^2}+\frac{5 g_{12}^2 g_{1c}^2}{12 \Delta _{12} \Delta _{1c} \mleft(\alpha _1+\Delta _{12}\mright) \mleft(\Delta _{1c}+\alpha _1\mright)}-\frac{5 g_{12}^2 g_{2c}^2}{6 \Delta _{2c}^2 \mleft(\alpha _2-\Delta _{12}\mright)^2}\\
&  + \frac{5 g_{1c}^2 g_{2c}^2}{3 \Delta _{2c}^2 \mleft(\alpha _{\text{c}}-\Delta _{1c}\mright)^2}-\frac{5 g_{1c}^2 g_{2c}^2}{6 \Delta _{2c}^2 \mleft(\Delta _{1c}+\alpha _1\mright)^2}-\frac{14 g_{1c}^2 g_{2c}^2}{3 \Delta _{1c}^2 \Delta _{2c}^2}-\frac{5 g_{12}^2 g_{2c}^2}{6 \Delta _{2c}^2 \mleft(\alpha _1+\Delta _{12}\mright)^2}-\frac{5 g_{12}^2 g_{2c}^2}{24 \Delta _{12}^2 \Delta _{2c}^2} -\frac{7 g_{2c}^4}{3 \Delta _{2c}^4} \\
&  -\frac{10 g_{1c}^2 g_{2c}^2}{3 \Delta _{1c} \Delta _{2c} \mleft(\Delta _{1c}-\alpha _{\text{c}}\mright) \mleft(\alpha _{\text{c}}-\Delta _{2c}\mright)}+\frac{5 g_{12}^2 g_{2c}^2}{12 \Delta _{12} \Delta _{2c} \mleft(\Delta _{12}-\alpha _2\mright) \mleft(\Delta _{2c}+\alpha _2\mright)}+\frac{5 g_{12}^2 g_{2c}^2}{4 \mleft(\alpha _2-\Delta _{12}\mright)^2 \mleft(\Delta _{2c}+\alpha _2\mright)^2}
\end{aligned} \mright) .
\end{align}

For clarity and brevity, we also show the resulting expressions up to second order, which are what one obtains when adding anharmonicity to the second-order expression in Eq.~(16) in the main text:
\begin{align}
\delta F^{\text{relaxation}} =& -\frac{2}{5}\Gamma_{1}^{q1} \mleft( 1 -\frac{g_{1c}^2}{\Delta _{1c}^2}-\frac{g_{12}^2}{\mleft(\alpha _2-\Delta _{12}\mright)^2}+\frac{g_{12}^2}{\mleft(\alpha _1+\Delta _{12}\mright)^2} \mright) \nonumber \\
&-\frac{2}{5} \Gamma_{1}^{q2}\mleft( 1 -\frac{g_{2c}^2}{\Delta _{2c}^2}+\frac{g_{12}^2}{\mleft(\alpha _2-\Delta _{12}\mright)^2}-\frac{g_{12}^2}{\mleft(\alpha _1+\Delta _{12}\mright)^2} \mright)\nonumber \\
& -\frac{2}{5}\Gamma_{1}^{c} \mleft( \frac{g_{1c}^2}{\Delta _{1c}^2}+\frac{g_{2c}^2}{\Delta _{2c}^2} \mright) ,
\end{align}
\begin{align}
\delta F^{\text{dephasing}} =& -\frac{2}{5}\Gamma_{\phi}^{q1} \mleft( 1 + \frac{g_{1c}^2}{2 \Delta _{1c}^2}+\frac{g_{12}^2}{2 \mleft(\alpha _2-\Delta _{12}\mright)^2}+\frac{9 g_{12}^2}{2 \mleft(\alpha _1+\Delta _{12}\mright)^2} \mright) \nonumber \\
&-\frac{2}{5} \Gamma_{\phi}^{q2}\mleft( 1 + \frac{g_{2c}^2}{2 \Delta _{2c}^2}+\frac{9 g_{12}^2}{2 \mleft(\alpha _2-\Delta _{12}\mright)^2}+\frac{g_{12}^2}{2 \mleft(\alpha _1+\Delta _{12}\mright)^2} \mright) \nonumber \\
& -\Gamma_{\phi}^{c} \mleft( \frac{g_{1c}^2}{\Delta _{1c}^2}+\frac{g_{2c}^2}{\Delta _{2c}^2}  \mright) .
\end{align}

\section{Numerical evaluation of the average gate fidelity in Eq. (1)}\label{sim-fid}

To evaluate the average gate fidelity in Eq.~(1) numerically, we use the generalized formulation of Ref.~\cite{wood2018}, which accounts for leakage outside the computational subspace and relaxes the trace-preserving assumption underlying Ref.~\cite{Nielsen2002}.
The average gate fidelity is given by
\begin{equation} \label{fid-process}
\bar{F} = \frac{d\,F_{\mathrm{pro}}(\mathcal{E}, g) + 1 - \Lambda_{\rm total}}{d + 1},
\end{equation}
where $d$ is the dimension of the computational subspace and $F_{\mathrm{pro}}(\mathcal{E},g)$ is the process fidelity, which measures how closely the applied map $\mathcal{E}$ matches the ideal gate $g$ and is defined as
\begin{equation}
F_{\mathrm{pro}}(\mathcal{E},g) = \frac{1}{d^2} \mathrm{Tr}\!\mleft[ S_{g}^\dagger\, S_\mathcal{E} \mright],
\end{equation}
where $S_g$ and $S_\mathcal{E}$ are the superoperator matrix representations of $g$ and $\mathcal{E}$ restricted to the computational subspace.
 The quantum map $\mathcal{E}$ is obtained by evolving a set of input states $\{|0\rangle, |1\rangle, |+x\rangle, |+y\rangle\}^{\otimes 2}$ spanning the computational subspace.
The leakage $\Lambda_{\rm total}$ quantifies population loss from the computational subspace and is given by
\begin{equation}
\Lambda_{\rm total} = 1 -  \mathrm{Tr}\!\mleft[ \hat{\mathds{1}}_{\rm cmp}\, \mathcal{E}\mleft(\frac{\mathds{1}_{\rm cmp}}{d}\mright)\mright],
\end{equation}
where $\hat{\mathds{1}}_{\rm cmp}$ denotes the identity operator on the computational subspace.

We emphasize that the goal of our numerical simulations is to extract the fidelity reduction arising specifically from decoherence processes, i.e., relaxation ($T_1$) and pure dephasing ($T_\phi$). However, Eq.~(1) in the main text and, consequently, \eqref{fid-process}, capture all sources of fidelity reduction. This includes not only decoherence, but also coherent errors such as leakage out of the computational subspace or imperfections in gate operations (e.g., inaccurate rotation angles in a CZ gate).
To distinguish these contributions, we define the coherent fidelity $F_{\rm coherent}$ as the fidelity obtained from \eqref{fid-process} when all decay rates are set to zero. This quantity therefore captures only coherent error mechanisms.
By subtracting this contribution from the total fidelity, we extract the reduction due solely to decoherence processes. In practice, this is done by taking the difference between the fidelity with finite decay rates and the fidelity in the absence of decay, $F_{\rm decoherence}= \bar{F} - F_{\rm coherent}$. The resulting quantity corresponds to the fidelity reduction due to decoherence, which can then be directly compared with the analytical results presented in this work.

\section{System parameters for adiabatic CZ gate simulations}\label{sim_detail}
The parameter values used in Eq.~(15) in the main text for simulations of the implementation of adiabatic CZ gates with tunable couplers, with results shown in Fig.~1 in the main text, are listed in Table~\ref{tab:sys_params}. 
\begin{table}[h!]
\centering
\begin{tabular}{lccccccc}
\hline \hline
$i$ & $\omega_i/2\pi$ & $\omega_{i,0}/2\pi$ & $\alpha_i/2\pi$ & $g_{ic,0}/2\pi$ & $g_{i1,0}/2\pi$ & $T_1$ ($\mu$s) & $T_\phi$ ($\mu$s) \\
\hline
1 & $4.65$ & --- & $-0.25$ & $0.070$ & --- & $100$ & $100$ \\
2 & $4.77$ & --- & $-0.25$ & $0.070$ & $-0.005$ & $100$ & $100$ \\
c & $4.023^*$ & $5.0$ & $-0.15$ & --- & $0.070$ & $20$ & $ 5^\dag$ \\
\hline \hline
\end{tabular}
\caption{Parameters used in the simulations of the adiabatic CZ gate. All frequency, anharmonicity, and coupling values are given in units of gigahertz. The effective capacitive coupling between elements $i$ and $j$ is frequency-dependent, modeled as $2g_{ij}= k_{ij}\sqrt{\omega_i\omega_j}$, where $k_{ij}$ is the static dimensionless coupling constant. The couplings $g_{ij,0}$ are evaluated at the zero-flux point ($\omega_{c,0}/2\pi = \SI{5.0}{\giga\hertz}$). \textsuperscript{*}Coupler idling frequency. \textsuperscript{\dag}Dephasing time evaluated at \SI{3}{\giga\hertz}.}
\label{tab:sys_params}
\end{table}

\end{document}